\documentclass{jfm}

\usepackage{graphicx,natbib,amssymb,amsmath,dashrule,color,psfrag,bm}
\usepackage{hyperref}
\hypersetup{
	colorlinks=true,
	linkcolor=black,
	citecolor=black
}

\newcommand\Ray{\mbox{\textit{Ra}}}  
\newcommand\Nus{\mbox{\textit{Nu}}}  
\newcommand\Ric{\mbox{\textit{Ri}}}  
\newcommand\Tay{\mbox{\textit{Ta}}}  
\newcommand\ie{i.e.\ }
\newcommand\cf{cf.\ }
\newcommand{\romd}{\mathrm{d}}
\DeclareMathOperator{\erf}{erf}

\shorttitle{BL changes at the edge of the ultimate regime in VC}
\shortauthor{C. S. Ng, A. Ooi, D. Lohse and D. Chung}
\title{Changes in the boundary-layer structure at the edge of the
	ultimate regime in vertical natural convection}
\author
{Chong Shen Ng$^1$
	\thanks{Email address for correspondence: chongn@unimelb.edu.au},\ns
	Andrew Ooi$^1$, Detlef Lohse$^2$, and Daniel Chung$^1$}

\affiliation{Department of Mechanical Engineering,
	The University of Melbourne, Victoria 3010, Australia\\[\affilskip]
	$^2$Physics of Fluids Group, Faculty of Science and Technology,
	J.\ M.\ Burgers Center for Fluid Dynamics and MESA+ Institute,
	\\University of Twente, 7500 AE Enschede, The Netherlands}

\date{?; revised ?; accepted ?. - To be entered by editorial office}

\begin{document}
\maketitle

\begin{abstract}

In thermal convection for very large Rayleigh numbers ($\Ray$), the thermal and 
viscous boundary layers are expected to undergo a transition from a classical 
state to an ultimate state. In the former state, the boundary layer thicknesses 
follow a laminar-like Prandtl--Blasius--Polhausen scaling, whereas in the 
latter, the boundary layers are turbulent with logarithmic corrections in the 
sense of Prandtl and von K{\'a}rm{\'a}n. Here, we report evidence of this 
transition via changes in the boundary-layer structure of vertical natural 
convection (VC), which is a buoyancy driven flow between differentially heated 
vertical walls. The numerical dataset spans $\Ray$-values from $10^5$ to $10^9$ 
and a constant Prandtl number value of $0.709$. For this $\Ray$ range, 
the VC flow has been previously found to exhibit classical state behaviour in a 
global sense. Yet, with increasing $\Ray$, we observe that near-wall 
higher-shear patches occupy increasingly larger fractions of the wall-areas, 
which suggest that the boundary layers are undergoing a transition from the 
classical state to the ultimate shear-dominated state. The presence of streaky 
structures--reminiscent of the near-wall streaks in canonical wall-bounded 
turbulence--further supports the notion of this transition. Within the 
higher-shear patches, conditionally averaged statistics yield a logarithmic 
variation in the local mean temperature profiles, in agreement with the log-law 
of the wall for mean temperature, and a $\Ray^{0.37}$ effective power-law 
scaling of the local Nusselt number. The scaling of the latter is consistent 
with the logarithmically corrected $1/2$-power law scaling predicted for 
ultimate thermal convection for very large $\Ray$. Collectively, the results 
from this study indicate that turbulent and laminar-like boundary layer coexist 
in VC at moderate to high $\Ray$ and this transition from the classical state to 
the ultimate state manifests as increasingly larger shear-dominated patches, 
consistent with the findings reported for Rayleigh--B{\'e}nard convection and 
Taylor--Couette flows.

\end{abstract}

\section{Introduction}

\begin{figure}
	\centering
	\centerline{\includegraphics{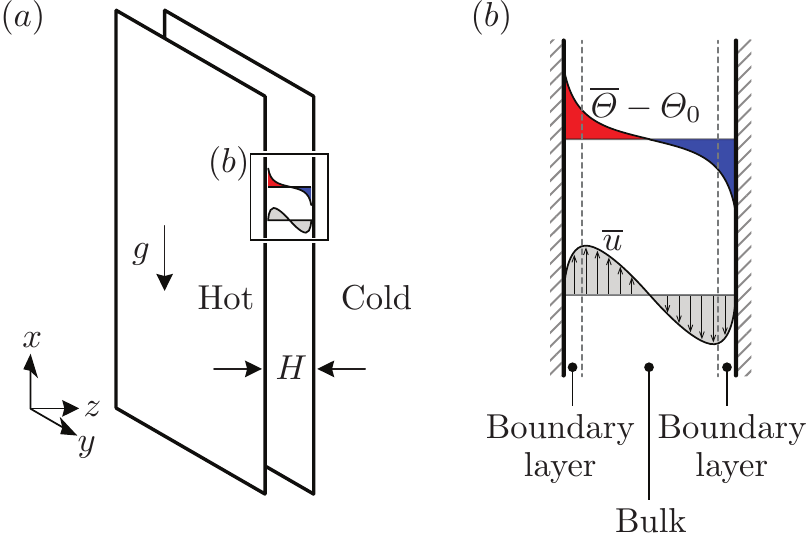}}
	\caption{\label{fig:VCSetup}($a$) Setup of vertical natural convection for the 
	present study. ($b$) Illustration of the mean temperature profile (top) and mean 
	streamwise velocity	profile (bottom). Both mean profiles are statistically 
	antisymmetric about the channel centreline. Also illustrated are the boundary 
	layers and bulk region. Note that in general, thermal and kinetic boundary 
	layers have different thicknesses.}
\end{figure}

Flows driven by thermal natural convection are ubiquitous in nature and in many 
engineering applications. In one of its many forms, a flow is sustained by simply 
applying a temperature difference at two opposing walls. For example when the walls 
are horizontal and the bottom wall is heated, gravity acts parallel to the heat 
flux and the setup is known as the classical Rayleigh--B{\'e}nard convection (RBC) 
\citep*{Ahlers+others.2009,Lohse+Xia.2010,Chilla+Schumacher.2012}. Alternatively, 
if heating and cooling are applied simultaneously at the bottom wall, the setup is 
referred as horizontal convection (HC) 
\citep*{Hughes+Griffiths.2008,Shishkina+Grossmann+Lohse.2016}. If the walls are 
instead vertical with one wall heated and the other cooled, gravity acts orthogonal 
to the heat flux and this setup is referred to as vertical natural convection 
(VC) \citep{Ng+Ooi+Lohse+Chung.2014,Shishkina.2016.momentum}. An example of this 
setup is shown in figure \ref{fig:VCSetup}. We note that in this configuration, the 
mean vertical pressure gradient is zero for VC. The transition between the extremes, 
RBC and VC, is continuous and \citet{Shishkina+Horn.2016} analysed thermal 
convection with some tilt angle $\alpha$ with $0^\circ \leqslant \alpha \leqslant 
90^\circ$ between the direction of gravity and the mean temperature gradient. 
Although $\alpha$ can be an input parameter, we consider only the cases for 
constant $\alpha$ (for VC, $\alpha = 0^\circ$). The two parameters that define all 
these setups are the Rayleigh number, $\Ray$, which is the dimensionless temperature 
difference, and Prandtl number, $\Pran$, which defines the fluid based on the ratio 
of kinematic viscosity to thermal diffusivity of the fluid. At the walls, viscous 
and thermal boundary layers form and are known to modulate the heat flux in the 
system, which is defined by the Nusselt number, $\Nus$. 

With increasing $\Ray$, the boundary layers in thermal convection become 
increasingly thinner and the flow is understood to undergo a transition from a 
`classical' regime, where the boundary layers follow a laminar-like 
Prandtl--Blasius--Pohlhausen (PBP) scaling, to an `ultimate' regime, where the 
boundary layers are turbulent in the sense of Prandtl and von K{\'a}rm{\'a}n 
\citep{Grossmann+Lohse.2011,Grossmann+Lohse.2012}. We are particularly interested 
to investigate the changes to the boundary-layer structure towards the ultimate 
regime and the accompanying $\Ray$-dependence of the Nusselt number and the 
Reynolds number $\Rey$. Our investigation is motivated by the fact that experimental 
studies in RBC have found a transition from the classical to the ultimate regime at 
a transitional Rayleigh number-value of approximately $5\times 10^{14}$
\citep{He+others.2012}. Accompanying the transition to the ultimate regime, the 
overall mean temperature profile exhibits an increasingly pronounced logarithmic 
dependence \citep{Ahlers+Others+Log.2012,Ahlers+Bodenschatz+He.2014}. In a 
two-dimensional numerical study at lower $\Ray$, the {\it local} mean temperature 
profile was also found to exhibit similar logarithmic variations where thermal 
plumes are emitted \citep{vanderPoel+coworkers.2015.logarithmic}. The 
presence of a local logarithmic-type mean temperature profile in RBC suggests that 
we can employ a criterion to distinguish flow regions in VC to compute local 
statistics, similar to the approach adopted by 
\cite{vanderPoel+coworkers.2015.logarithmic}. In addition, the big and unique 
advantage of VC is that a well-defined mean flow exists which allows for a longer 
time sampling of our data. In \cite{Ng+Ooi+Lohse+Chung.2014}, we have also shown 
that aspects of the Grossmann--Lohse theory 
\citep{Grossmann+Lohse.2000,Grossmann+Lohse.2001,Grossmann+Lohse.2002,Grossmann+Lohse.2004}
(hereafter GL theory), which provided physical arguments for the scaling relations 
in horizontal thermal convection, \ie RBC, can also be extended to VC. Therefore, 
the present study makes contact with current efforts to understand the changes to 
the boundary layer structure in the transition from the classical to the ultimate 
regime of thermal convection.

The paper is organized as follows: In \S\,\ref{sec:GovEquations}, we first give the 
underlying dynamical equations which are numerically solved. From a visual 
assessment of the near-wall velocity and temperature fields (\S\,\ref{sec:FlowVis}), 
we identify two regions that are increasingly distinguishable at higher $\Ray$, 
namely a streaky higher-shear flow region and a non-streaky lower-shear flow region. 
Motivated by the visual changes in the near-wall flow features, we define a 
flux-based Richardson number, $|\Ric_f|$ (\S\,\ref{sec:RichardsonNo}), which is
inspired by the classical definition \citep[see for example in Chapter 5 
of][]{Turner.1979}. A criterion is selected based on $|\Ric_f|$ in order to 
separate the dynamics of the streaky and non-streaky regions. Conveniently, the 
streaky higher-shear flow regions are well-correlated with low-$|\Ric_f|$ regions, 
and vice versa. When we quantify the wall-area fraction occupied by the streaky 
higher-shear regions (\S\,\ref{sec:TurbFraction}), we find that the wall-area 
fraction increases with increasing $\Ray$. We then employ a conditional averaging 
procedure to compute the local mean statistics of the two regions 
(\S\,\ref{sec:MeanStats}) as well as the local scaling of $\Nus$ and $\Rey$ 
(\S\,\ref{sec:NuReScaling}). Finally, we compare the near-wall structures in VC 
with the near-wall streaks in pressure-driven channel flow, the latter of which are 
unique and robust features of the buffer region in wall-bounded turbulence. The 
comparison is aided by the analysis of the 	one-dimensional premultiplied spectra 
of streamwise velocity (\S\,\ref{sec:CondSpectra}). The paper ends with a summary 
and conclusions (\S\,\ref{sec:Conclusion}).

\section{Dynamical equations and boundary conditions}\label{sec:GovEquations}
In this study, we analyse the dataset for VC for Rayleigh numbers ranging from 
$10^5$ to $10^9$ at constant Prandtl number value of 0.709. The numerical setup for 
VC is the same as that employed by \cite{Ng+Ooi+Lohse+Chung.2014}, see figure 
\ref{fig:VCSetup}, \ie we consider a buoyancy driven flow between two vertical 
walls with the left wall heated and the right wall cooled. The temperatures at 
the wall are kept uniform and are denoted $T_h$ and $T_c$, respectively. Thus, the 
temperature difference of the walls $\Delta = T_h - T_c$ establishes a heat flux 
which acts horizontally across a wall-separation distance $H$. The governing 
continuity, momentum and energy equations for the velocity field 
$\boldsymbol{u}(\boldsymbol{x},t)$ and the temperature field 
$\varTheta(\boldsymbol{x},t)$ are respectively given by,
\begin{subeqnarray}
\bnabla\bcdot\boldsymbol{u} &=& 0,  \label{eqn:NonDimContEqn}  \\ 
\dfrac{\partial \boldsymbol{u}}{\partial t} + \boldsymbol{u} \bcdot \bnabla 
\boldsymbol{u} &=& -\frac{1}{\rho_0} \nabla p + \hat{\mathbf{e}}_x g 
\beta 
(\varTheta-\varTheta_0) + \nu \nabla^2 \boldsymbol{u}, \label{eqn:NonDimMomEqn}\\	
\dfrac{\partial \varTheta}{\partial t} + \boldsymbol{u} \bcdot \bnabla 
\varTheta
&=&	\kappa \nabla^2 \varTheta, \label{eqn:NonDimTempEqn}
\end{subeqnarray}
\returnthesubequation
where $\hat{\mathbf{e}}_x$ is the unit vector in the $x$-direction. The 
pressure field is denoted by $p(\boldsymbol{x},t)$. We define $\varTheta_0 = 
(T_h + T_c)/2$ as the reference temperature and $g$ as the gravitational 
acceleration. We also specify $\beta$ as the coefficient of thermal expansion, 
$\nu$ as the kinematic viscosity and $\kappa$ as the thermal diffusivity, all 
assumed to be independent of temperature. The Rayleigh and Prandtl numbers are then 
respectively defined as
\begin{subeqnarray}
	\Ray \equiv \dfrac{g\beta\Delta H^3}{\nu\kappa},\quad 
	\Pran \equiv \dfrac{\nu}{\kappa}.
\end{subeqnarray}
\returnthesubequation
The coordinate system $x$, $y$ and $z$ refers to the streamwise (opposing 
gravity), spanwise and wall-normal directions. The no-slip and no-penetration 
boundary conditions are imposed on the velocity at the walls. Periodic boundary 
conditions are imposed on $\boldsymbol{u}$, $p$ and $\varTheta$ in the $x$- 
and $y$-directions. In addition, we denote time- and $xy$-plane-averaged 
quantities with an overbar, and the corresponding fluctuating part with a prime, 
\eg the mean streamwise velocity component is defined by $\overline{u}=u-u^\prime$ 
and the mean temperature is defined by $\overline{\varTheta} = \varTheta - 
\varTheta^\prime$.

Equations (\ref{eqn:NonDimContEqn}$a$--$c$) are numerically solved in the 
domain-size $L_x \times L_y \times L_z = 8H \times 4H \times H$. The present 
simulations employ smaller periodic-domain sizes (two-thirds in each periodic 
direction) than other DNS studies 
\citep[\eg][]{Versteegh1999,Pallares+Vernet+Ferre+Grau.2010} but are chosen in order 
to resolve the near-wall region at high $\Ray$. Comparison with previous DNS studies 
showed good agreement for the mean and second-order statistics and other simulation 
parameters have been previously reported in \cite{Ng+Ooi+Lohse+Chung.2014}. All 
of the analyses in this study are based on statistics that are averaged from both 
halves of the channel, taking the statistical antisymmetry (about the centreline) of 
the mean profiles into account.
\begin{figure}
	\centering
	\centerline{\includegraphics{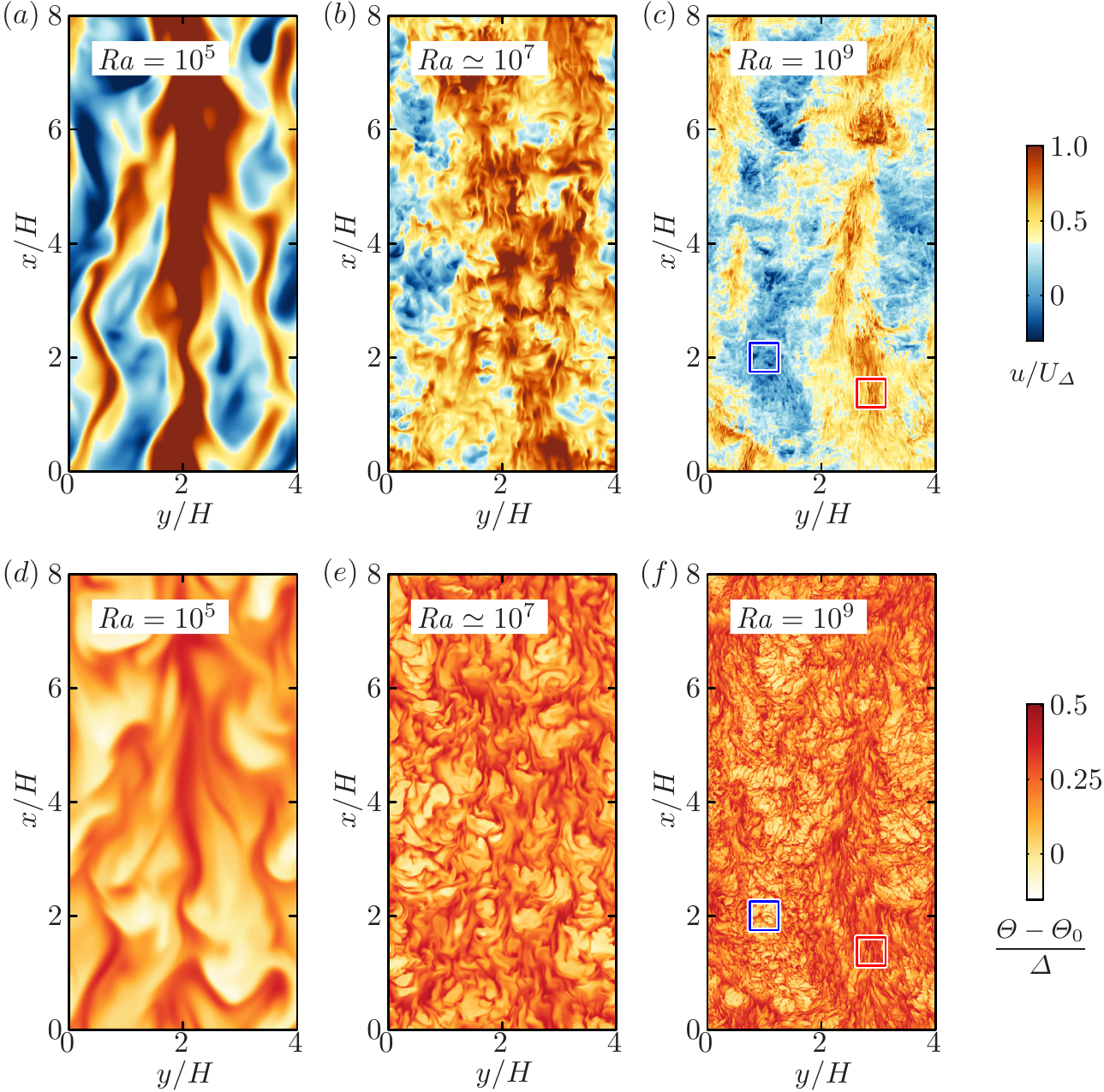}}
	\caption{\label{fig:VelTempSlice}($a$,$b$,$c$) Wall-parallel planes of 
	streamwise velocity, and ($d$,$e$,$f$) temperature for $\Ray\simeq10^5$, $10^7$ 
	and $10^9$. The planes are extracted close to the hotter wall at $z^+ \approx 
	15$. Gravity is acting downwards. The non-streaky lower-shear regions are 
	indicated by the blue boxes whereas the streaky higher-shear regions are 
	indicated by the red boxes (see magnified views in figure 
	\ref{fig:VelTempSliceZoom}). Here, $U_\Delta\equiv \sqrt{g\beta\Delta H}$ the 
	free-fall velocity.}
\end{figure}

\section{Flow visualisations}\label{sec:FlowVis}
Figure \ref{fig:VelTempSlice} shows the wall-parallel planes of velocity and 
temperature for $\Ray\simeq10^5$, $10^7$ and $10^9$. The planes were extracted from 
the three-dimensional volumes for velocity and temperature near the hotter wall at 
the wall-normal location equivalent to $z^+ \equiv z/\delta_\nu \approx 15$, where
$\delta_\nu \equiv \nu/u_\tau$ is the viscous length scale and $u_\tau \equiv 
\sqrt{\nu\,\romd \overline{u}/\romd z|_w}$ is the friction velocity scale. Note that 
we use the $^+$ symbol to denote normalisation in viscous wall units. In figures 
\ref{fig:VelTempSlice}($a$--$c$), the velocity fields are coloured from blue to red 
to denote low- and high-speed regions (or equivalently, lower- and higher-shear 
regions), respectively, whereas in figures \ref{fig:VelTempSlice}($d$--$f$) the 
temperature fields are coloured from white to red to denote the temperature 
variations from cooler regions with a temperature around $\varTheta_0$ to hotter 
regions with a temperature around $T_h$.
\begin{figure}
	\centering
	\centerline{\includegraphics{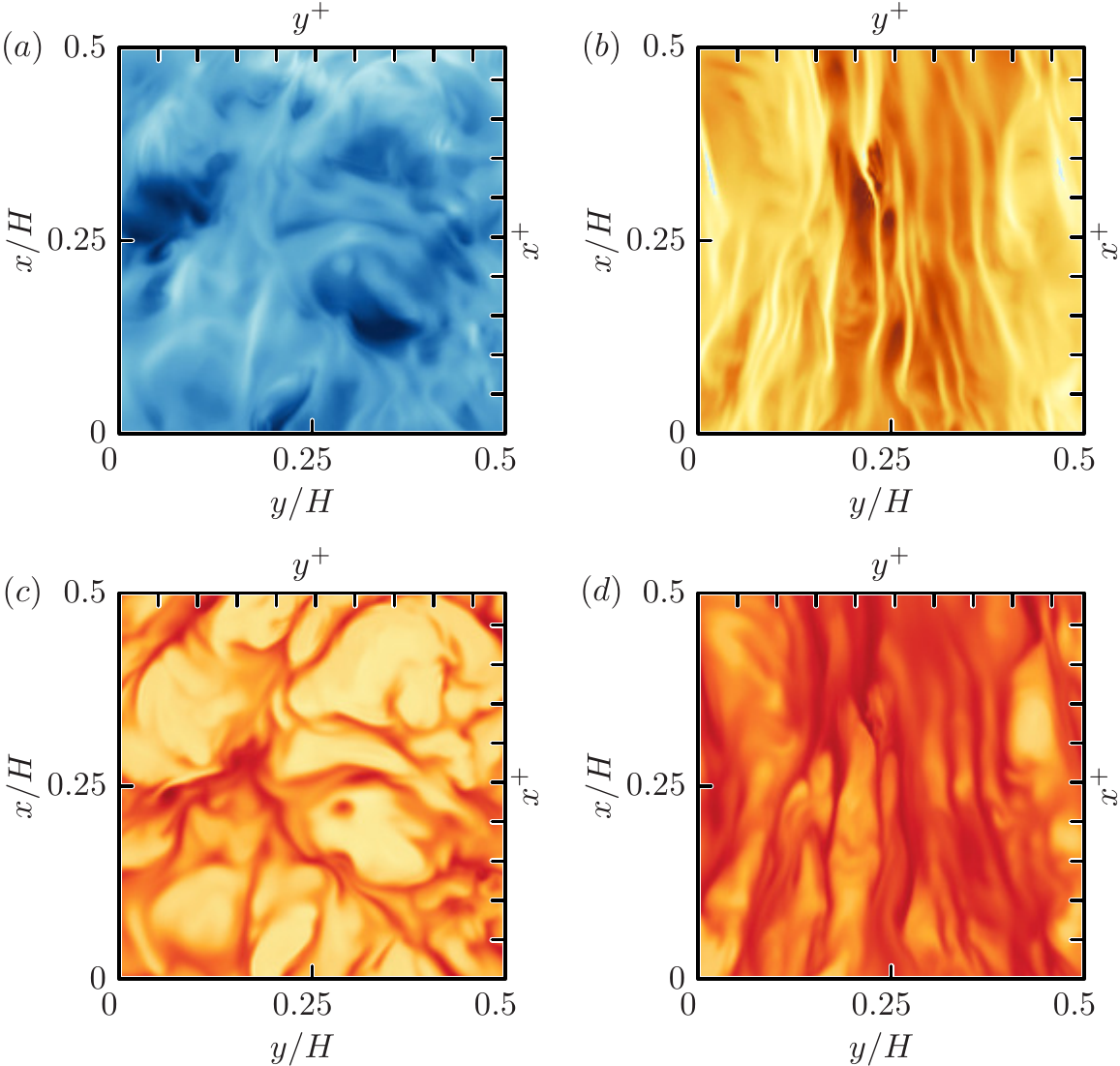}}
	\caption{\label{fig:VelTempSliceZoom}Magnified wall-parallel planes of 
	streamwise velocity ($a$,$b$) and temperature ($c$,$d$), respectively 
	reproduced from the blue and red boxes from figure	
	\ref{fig:VelTempSlice}($c$,$f$) for $\Ray=10^9$. In ($a$,$c$) we observe 
	non-streaky lower-shear regions (from the blue boxes), whereas in ($b$,$d$) we 
	observe streaky higher-shear regions (from the red boxes), both with 
	characteristic spanwise spacings of 100 to 200 viscous units. For reference, the 
	axes at the top and right edges denote spacings of 100 viscous units. The 
	colormaps are the same as in figure \ref{fig:VelTempSlice}.}
\end{figure}
Not surprisingly, from figure \ref{fig:VelTempSlice}, the structures in the velocity 
and temperature fields are observed to become increasingly finer at higher $\Ray$. 
The finer structures are particularly prominent at $\Ray=10^9$ (figures 
\ref{fig:VelTempSlice}$c$ and $f$). Coincident with the higher-shear regions, the 
finer structures appear streaky and are markedly different to the structures in the 
lower-shear regions as well as the structures in the visualisations at 
lower-$\Ray$. To focus on the differences, magnified views of the higher- and 
lower-shear regions are plotted for $\Ray=10^9$ in figure \ref{fig:VelTempSliceZoom}.

On closer inspection of the magnified views (figure \ref{fig:VelTempSliceZoom}), the 
streaky structures appear reminiscent of the near-wall streaks, which are 
well-documented flow features in canonical turbulent boundary layers 
\citep{Hutchins+Marusic_MeanderingFeatures.2007,Marusic+Others.2010,Smits+McKeon+Marusic.2011}.
In the magnified view, the spanwise spacing of the streaky structures is 
approximately 100--200 viscous units, consistent with the spacing of the near-wall 
streaks found by \cite{Kline+Others.1967}. The similarity between the streaky
structures in VC at higher $\Ray$ and in canonical turbulent boundary layers 
suggests that locally, the boundary layers in VC are much more strongly animated by 
a large-scale `wind', perhaps an indication of an incipient turbulent boundary 
layer. Additionally, this incipient behaviour appears to manifest 
as increasingly larger streaky regions, as observed in the plane visualisations in 
figure \ref{fig:VelTempSlice}. Therefore, a reasonable starting point would be to 
seek a quantity that is able to distinguish between the two regions and that would 
also enable us to study the corresponding local scaling.

\section{Identifying different flow regions}\label{sec:RichardsonNo}

In order to quantitatively distinguish the two regions shown in the magnified views 
in figure \ref{fig:VelTempSlice}, we define a flux-based Richardson number at the 
wall-normal location $z^+ = 15$ (\ie\,just outside the viscous sublayer),
\begin{equation}
\Ric_f \equiv \dfrac{z(z^+ = 15)}{L} = \dfrac{15}{L^+}, \label{eqn:Richardson}
\end{equation}
where $L \equiv -u_\tau^3/(g\beta f_w)$ is the Obukhov length and $f_w \equiv 
-\kappa (\romd \overline{\varTheta}/\romd z)|_w$ is the wall heat flux and $u_\tau$ 
is the friction velocity scale defined in \S\,\ref{sec:FlowVis}. The Obukhov 
length is defined such that when $f_w<0$, then $L>0$ 
\citep[\S\,7.2][]{Monin2007statisticalVol1}. Since $f_w>0$ for VC, then $L<0$ and 
the sign of $\Ric_f$ becomes negative for VC. Therefore, to assist interpretation, 
we will use $\Ric_f$ mainly modulo-wise, $|\Ric_f|$.

We emphasise that our flux-based Richardson number definition in 
(\ref{eqn:Richardson}) is a {\it local} quantity (in the plane $z^+ = 15$) to 
measure shear-dominated areas versus buoyancy dominated areas and is merely an 
analogy that is inspired by the classical definition \citep[see for example Chapter 
5 in][]{Turner.1979}. Equation (\ref{eqn:Richardson}) is notionally similar to the 
Obukhov stability parameter, which is a dimensionless height typically adopted in 
the study of atmospheric surface layer turbulence \citep{Businger+Others.1971}. Our
definition is therefore far from perfect since the vertical heat flux in VC is a 
response to the horizontal heat flux imposed on the system, whereas in the 
classical definition, both the imposed and responding heat fluxes act in the same 
(vertical) direction. Nevertheless, we take advantage of (\ref{eqn:Richardson}) in 
our study simply as a convenient parameter which is able to quantify the dominance 
of buoyancy relative to shear. It is also for this reason that we prefer to use 
(\ref{eqn:Richardson}) to distinguish high-speed regions instead of other 
quantities, such as the shear Reynolds number \citep[see for 
example][]{Scheel+Schumacher.2016}. By defining (\ref{eqn:Richardson}), we also 
make contact with recent efforts to distinguish `wind-sheared' regions in flows that 
are analogous to VC, for example, in Taylor--Couette flows 
\citep{Ostilla+Coworkers.2014} and RBC \citep{vanderPoel+coworkers.2015.logarithmic}.
\begin{figure}
	\def\drawline#1#2{\raise 2.5pt\vbox{\hrule width #1pt height #2pt}}
	\centering
	\centerline{\includegraphics{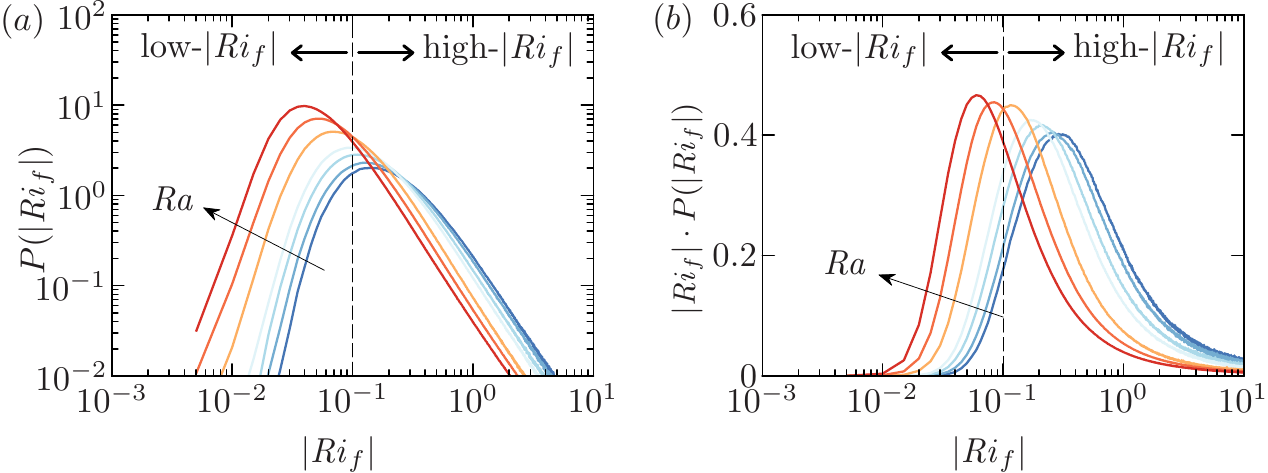}}
	\caption{\label{fig:RiPDF}($a$) Probability distribution functions of 
	$|\Ric_f|$, and ($b$) the premultiplied form defined in equation 
	(\ref{eqn:PremultiplyPDF}). The curves are coloured blue to red to represent 
	increasing $\Ray$. The low- and high-$|\Ric_f|$ ranges are demarcated			
	by $|\Ric_f| = 0.1$ (dashed line). The premultiplied PDFs in ($b$) are 
	increasingly weighted towards lower-$|\Ric_f|$ with increasing $\Ray$.}
\end{figure}

We now turn to the probability distributions functions (PDFs) of $|\Ric_f|$ (figure 
\ref{fig:RiPDF}$a$) to determine a suitable threshold value that can distinguish 
the 
high-speed and  low-heat flux regions. To visually aid the interpretation on a 
doubly logarithmic plot, we premultiply the PDFs with $|\Ric_f|$ so that the area 
of a semi-logarithmic plot represents the probability, \ie
\begin{equation}
\int_{0}^{+\infty} P(|\Ric_f|)\,\romd |\Ric_f| = 1 = \int_{0}^{+\infty} 
|\Ric_f|\cdot P(|\Ric_f|)\,\romd (\log |\Ric_f|). \label{eqn:PremultiplyPDF}
\end{equation}
The integrand on the left-hand-side of (\ref{eqn:PremultiplyPDF}) is plotted in 
figure \ref{fig:RiPDF}($a$) and the integrand on the right-hand-side of 
(\ref{eqn:PremultiplyPDF}) is plotted in figure \ref{fig:RiPDF}($b$). Therefore, in 
the representation of figure \ref{fig:RiPDF}($b$), the area beneath the 
PDF curve represents equal probability of $|\Ric_f|$, which is not the case 
in figure \ref{fig:RiPDF}($a$). The curves are coloured from blue to red 
to represent increasing $\Ray$. We further define a low-$|\Ric_f|$ as $0<|\Ric_f| 
\leqslant 0.1$ and high-$|\Ric_f|$ as $|\Ric_f| > 0.1$. From figure 
\ref{fig:RiPDF}($b$), we observe that with increasing $\Ray$, there is an increase 
in the areas under the premultipied curves which correspond to the low-$|\Ric_f|$. 
This increase in area suggests that the distributions of $|\Ric_f|$ are increasingly 
weighted towards the lower-$|\Ric_f|$ values with increasing $\Ray$. As a further 
example, the value of the mode of the Richardson number $|\Ric_{f,m}|$ in figure 
\ref{fig:RiPDF}($a$) decreases as $|\Ric_{f,m}| \approx 0.99 \Ray^{-0.16}$ from a 
least-squares fit to a single power law.

\begin{figure}
	\def\drawline#1#2{\raise 2.5pt\vbox{\hrule width #1pt height #2pt}}
	\centering
	\centerline{\includegraphics{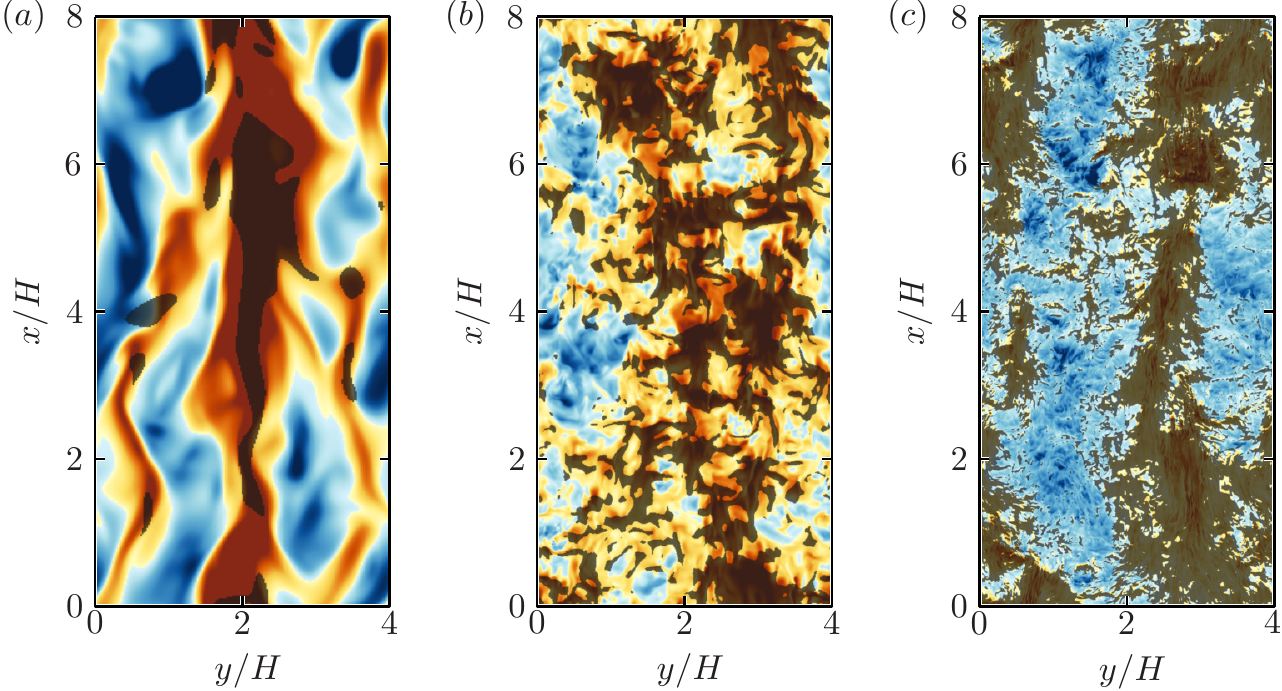}}
	\caption{\label{fig:RichardsonSlice}Same wall-parallel planes of streamwise 
	velocity as in figures \ref{fig:VelTempSlice}($a$--$c$), but now overlayed with 
	grey contours of $0 < |\Ric_f| \leqslant 0.1$. The $\Ray$-values are ($a$) 
	$10^5$,	($b$) $2\times 10^7$ and ($c$) $10^9$. With increasing $\Ray$, the grey 
	contours are increasingly correlated with the streaky higher-shear regions and 
	occupy increasingly larger wall-area fractions. Gravity	is acting downwards.}
\end{figure}
To explain the trend observed in the premultiplied PDFs, we refer to a visual  
example of low-$|\Ric_f|$ regions which are shown as grey-filled patches in figure 
\ref{fig:RichardsonSlice}. The grey-filled patches are overlayed on the 
plane-parallel visualisations of streamwise velocity for three $\Ray$-values, which 
are reproduced from figure \ref{fig:VelTempSlice}. At the highest $\Ray$ (figure 
\ref{fig:RichardsonSlice}$c$), we observe that the low-$|\Ric_f|$ regions are 
well-correlated with the streaky regions in figure \ref{fig:VelTempSlice}($c$). In 
figure \ref{fig:RichardsonSlice}, the degree of correlation also increases with 
increasing $\Ray$ and is matched by an increasingly larger wall-area coverage of 
low-$|\Ric_f|$ regions. The improved correlations and increasing wall-area coverage 
of the low-$|\Ric_f|$ regions are consistent with the increasingly weighted trends 
of the premultiplied PDFs of lower-$|\Ric_f|$ values with increasing $\Ray$, seen in 
figure \ref{fig:RiPDF}($b$). These well-matched behaviours of low-$|\Ric_f|$ 
distributions and the streaky regions imply that it may be possible to quantify 
shear-dominated regions from a carefully selected low-$|\Ric_f|$ value.

As a starting point, we choose the range $0 < |\Ric_f| \leqslant 0.1$ as the 
criteria for identifying shear-dominated regions in our study. If a lower or higher 
upper limit is selected, the corresponding wall-areas of the local flow that 
meet the criteria are smaller and larger, respectively. (A brief discussion about 
the sensitivity of wall-areas to the upper limit is given in Appendix 
\ref{sec:AppendixA}.) Not surprisingly, the choice of the upper limit has an 
influence on the local effective scaling. However, we will show in 
\S\,\ref{sec:NuReScaling} that for a sufficiently small $|\Ric_f|$ threshold, the 
effective local scaling exponent of the Nusselt number is unequivocally higher than 
the global effective scaling exponent and therefore, does not alter the conclusions 
of this study.

\section{Low-$|\Ric_f|$ area fraction}\label{sec:TurbFraction}
\begin{figure}
	\centering
	\centerline{\includegraphics{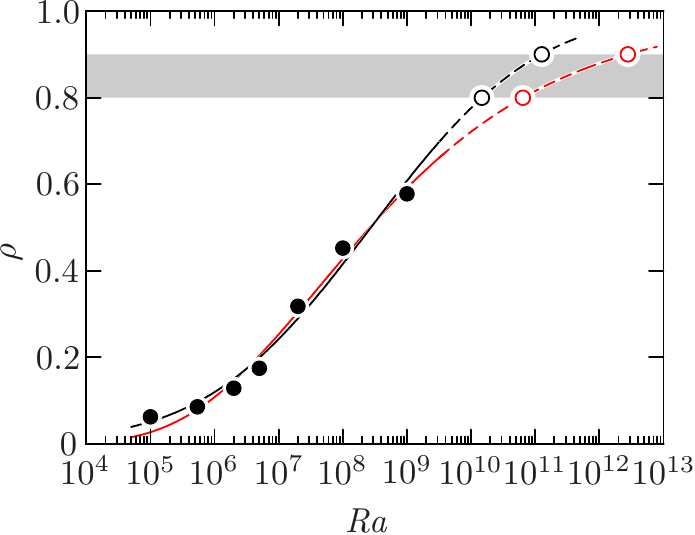}}
	\vspace*{0.5em}
	\caption{\label{fig:TurbFraction}Plot of $\rho$ the area fraction of 
	higher-shear low-$|\Ric_f|$ regions (solid circles) versus $\Ray$. $\rho$ is 
	defined in equation (\ref{eqn:TurbFraction}). The solid black curve is the 
	empirical fit to $\rho_{fit,1}$, see equation 	
	(\ref{eqn:TurbFractionCompositeFit}), and the solid red curve is the empirical 
	fit to $\rho_{fit,2}$, see equation (\ref{eqn:TurbFractionGompertzFit}). From an 
	extrapolation to $\rho = 0.8$ and $0.9$, we obtain $\Ray \approx 
	1.5\times10^{10}$ and $\approx 1.2\times10^{11}$, respectively, for 	
	$\rho_{fit,1}$ (black open circles) and $\Ray \approx 6.5\times 10^{10}$ and 
	$\approx 2.8\times 10^{12}$, respectively, for $\rho_{fit,2}$ (red open 	
	circles).}
\end{figure}

Using the flux-Richardson number criteria defined in \S\,\ref{sec:RichardsonNo}, we 
quantify the wall-areas occupied by the low-$|\Ric_f|$ regions for our $\Ray$-range. 
These wall-areas correspond to shear-dominated regions and we quantify
the area fraction $\rho(\Ray)$ according to
\begin{equation}
\rho(\Ray) \equiv \dfrac{\int_{0}^{L_y}\int_{0}^{L_x} I_{0 < |\Ric_f| \leqslant 
0.1}\,\romd x\,\romd y}{\int_{0}^{L_y}\int_{0}^{L_x}\,\romd x\,\romd y}, 
\label{eqn:TurbFraction}
\end{equation}
where $I$ is an indicator function, being 1 for wall-areas where $0 < |\Ric_f| 
\leqslant 0.1$ (low-$|\Ric_f|$) and being 0 where $|\Ric_f| > 0.1$ 
(high-$|\Ric_f|$). Therefore, the numerator represents the streamwise-spanwise 
wall-area influenced by low-$|\Ric_f|$ flow and the denominator represents the total 
wall-area of our setup, which is $L_x\times L_y =32H^2$. Equation 
(\ref{eqn:TurbFraction}) is similar to the definition used to quantify the spatial 
intermittency of strong shear regions in RBC \citep[\eg][]{Scheel+Schumacher.2016}. 
The results for the area fractions $\rho(\Ray)$ are shown in figure 
\ref{fig:TurbFraction} as filled circles.

In figure \ref{fig:TurbFraction}, the area fractions for the low-$|\Ric_f|$ regions 
range from $\rho \approx 0.062$ for $\Ray = 10^5$ to $\rho \approx 0.58$ for $\Ray 
= 10^9$. Visually, the trend of the area fractions appears sigmoidal and 
therefore, using a fit to an error function, the trend for $\rho(\Ray)$ can be 
approximated by
\begin{equation}
\rho_{fit,1} = 0.49 \erf\left[0.35\log_{10}(\Ray) - 2.95\right] + 0.5. 
\label{eqn:TurbFractionCompositeFit}
\end{equation}
Alternatively, the trend for $\rho(\Ray)$ can also be approximated by a fit to the 
Gompertz function
\begin{equation}
\rho_{fit,2} = 0.99\exp \{-2.21\exp\left[-0.48\log_{10}(\Ray) + 2.91 \right]\}, 
\label{eqn:TurbFractionGompertzFit}
\end{equation}
which is an asymmetric sigmoid function. In contrast to 
(\ref{eqn:TurbFractionCompositeFit}), equation (\ref{eqn:TurbFractionGompertzFit}) 
captures a lower inflection point in the trend of $\rho$, which occurs at $\Ray = 
2\times 10^7$ in figure \ref{fig:TurbFraction}. However, equation 
(\ref{eqn:TurbFractionGompertzFit}) does not capture the trend of $\rho$ for 
$\Ray\lesssim 5.4\times 10^5$ compared to (\ref{eqn:TurbFractionCompositeFit}). 
We note that both curve fits are only statistical approximations and there are 
inherent risks involved when extrapolating results to higher $\Ray$. Therefore, in 
the interest of obtaining a conservative estimate, we include both $\rho_{fit,1}$ 
and $\rho_{fit,2}$ in our subsequent discussions.

From extrapolations of (\ref{eqn:TurbFractionCompositeFit}) and 
(\ref{eqn:TurbFractionGompertzFit}), approximately $80\%$ to $90\%$ of the 
near-wall region are presumed to be dominated by streaks between $\Ray \approx 
1.5\times10^{10}$ to $1.2\times10^{11}$, based on 
(\ref{eqn:TurbFractionCompositeFit}), or $\Ray \approx 6.5\times 10^{10} $ to 
$2.8\times 10^{12}$, based on (\ref{eqn:TurbFractionGompertzFit}). Overall, the 
trend of $\rho(\Ray)$ is increasing and indicates that higher-shear regions occupy 
increasingly larger wall areas at higher $\Ray$.

Our results are consistent with the notion that for the present $\Ray$ range, the 
boundary layers in VC can be considered to be in a transitional state. As the flow 
approaches the fully turbulent, or ultimate regime, the relative size of the 
shear-dominated regions grow until the regions fully occupy the wall 
\citep{vanderPoel+coworkers.2015.logarithmic}.

\section{Mean profiles and conditional mean profiles}\label{sec:MeanStats}

\begin{figure}
	\def\drawline#1#2{\raise 2.5pt\vbox{\hrule width #1pt height #2pt}}
	\centering
	\centerline{\includegraphics{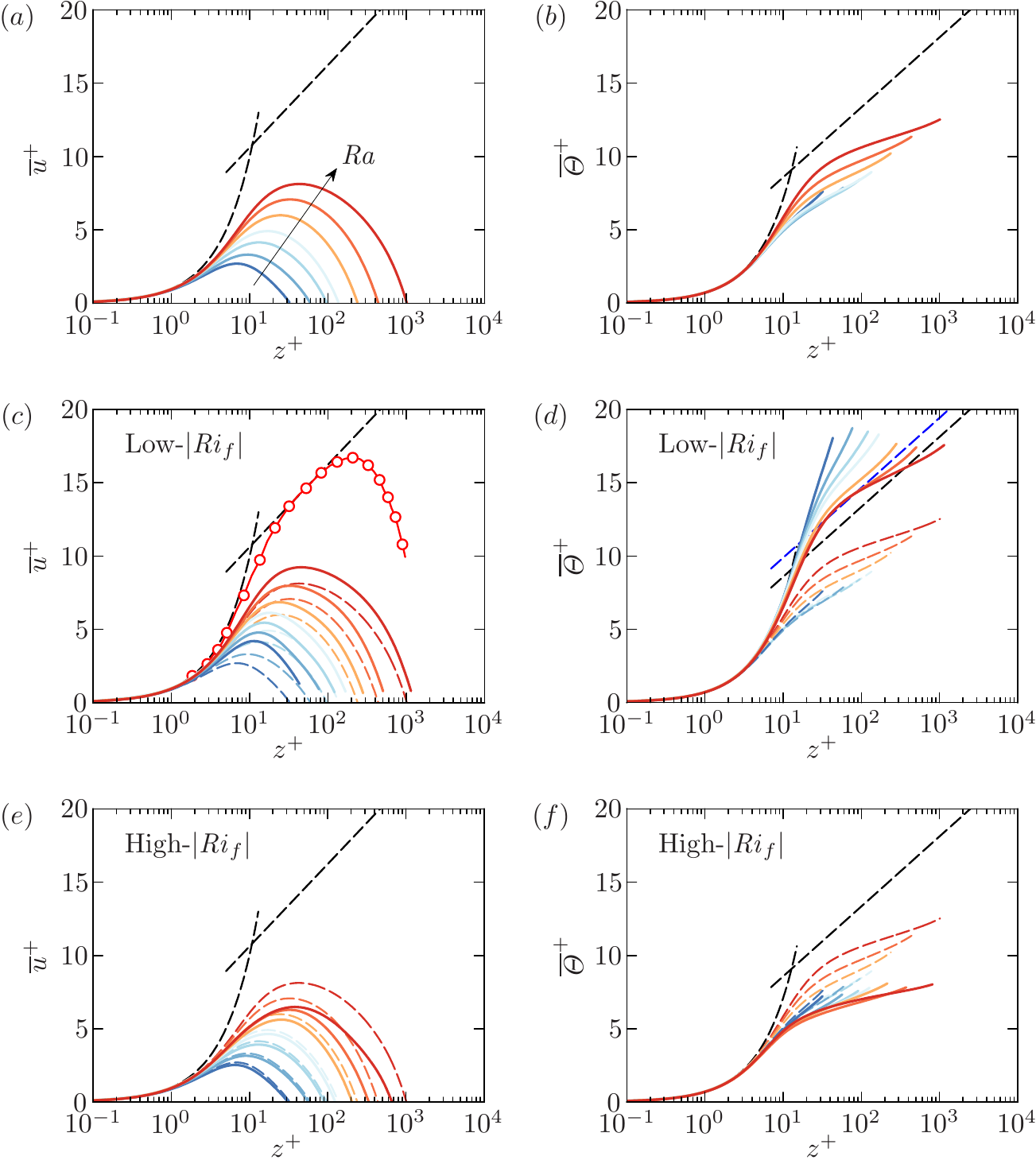}}
	\caption{\label{fig:MeanVelTemp}($a$) Overall mean velocity profiles, and ($b$) 
	overall mean temperature profiles, coloured from blue to red to represent 
	increasing $\Ray$. The overall mean profiles are repeated in ($c$) to ($f$) as 
	dashed curves. ($c$,$d$) Conditional mean profiles in low-$|\Ric_f|$ ($0 < 
	|\Ric_f| \leqslant  0.1$) regions. ($e$,$f$) Conditional mean profiles in 
	high-$|\Ric_f|$ ($|\Ric_f| >  0.1$) regions. The black straight dashed lines 
	correspond to the log-laws of the walls for velocity and temperature, equation 
	(\ref{eqn:LogLaws}), with $\kappa_\nu = 0.41$, $A = 5.0$, $\Pran_t = 0.85$ and 
	$B = 3.8$. In ($c$), the red curve with open symbols correspond to 
	the mean streamwise velocity from the wall-jet flow \citep[reproduced from 
	figure 13,][]{Eriksson+Others.1998}. In ($d$), the blue straight dashed line 
	corresponds to equation (\ref{eqn:LogLaws}$b$) with $B=5.1$.}
\end{figure}

The overall mean profiles for streamwise velocity and temperature are shown in 
figures \ref{fig:MeanVelTemp}($a$) and ($b$), where the profiles are coloured from 
blue to red to indicate increasing $\Ray$. From both figures, we conclude that 
there are no log-profiles in the mean statistics when scaled in wall units. We 
highlight the absence of the log-profiles in the overall mean profiles by including 
in figure \ref{fig:MeanVelTemp}($a$) and ($b$) the log-laws of the wall, which are 
expected when the flow is fully turbulent in the sense of Prandtl and von 
K{\'a}rm{\'a}n. The log-laws of the wall for streamwise velocity and temperature 
may be written as,
\begin{subeqnarray}\label{eqn:LogLaws}
	\gdef\thesubequation{\theequation \mbox{\textit{a}},\textit{b}}
	\overline{u}^+ = \dfrac{1}{\kappa_v}\log(z^+) + A \qquad\text{and}\qquad
	\overline{\varTheta}^+ = \dfrac{\Pran_t}{\kappa_v}\log(z^+) + B,
\end{subeqnarray}
\returnthesubequation
\citep[\cf][]{Yaglom.1979}, where $\overline{u}^+ \equiv \overline{u}/u_\tau$ and
$\overline{\varTheta}^+ \equiv (\overline{\varTheta}-\varTheta_h)/\theta_\tau$. 
Equations (\ref{eqn:LogLaws}) are plotted as black dashed lines in figure 
\ref{fig:MeanVelTemp}. For the purposes of our study, we choose to adopt the same 
constants as \cite{Yaglom.1979}, \ie the von K{\'a}rm{\'a}n constant $\kappa_v = 
0.41$, the turbulent Prandtl number $\Pran_t=0.85$, $A=5.0$ and $B=3.8$. We also 
employ the temperature scale $\theta_\tau \equiv -f_w/u_\tau$. Near the wall, the 
mean profiles are linear and obey $\overline{u}^+ = z^+$ and $\overline{\varTheta}^+ 
= \Pran\,z^+$.

To assess the mean profiles in the streaky and high-shear regions only, we compute 
the conditional mean profiles within the low-$|\Ric_f|$ regions according to the 
conditional averaging procedure
\begin{equation}
\overline{(\cdot)}_l(z) = \dfrac{\int_{0}^{L_y}\int_{0}^{L_x} I_{0 < |\Ric_f| 
\leqslant 0.1}(\cdot)(\boldsymbol{x},t)\,\romd x\,\romd 
y}{\int_{0}^{L_y}\int_{0}^{L_x}I_{0 < |\Ric_f| \leqslant 0.1}\,\romd x\,\romd y},
\label{eqn:ConditionalAveProcess}
\end{equation}
where the subscript $l$ distinguishes the quantities $(\cdot)$ in the low-$|\Ric_f|$ 
regions and $I$ is the indicator function previously defined in 
(\ref{eqn:TurbFraction}). The conditional mean profiles for the high-$|\Ric_f|$ 
regions, denoted by a subscript $h$, follow the similar definition as 
(\ref{eqn:ConditionalAveProcess}) but with the indicator function criteria 
$I_{|\Ric_f| > 0.1}$. Thus, the conditional mean and variance profiles are related 
to the area fraction $\rho$ by
\begin{subeqnarray}\label{eqn:ConditionalAveDecomposed}
	\overline{u}(z) &=& \rho \overline{u}_l + (1-\rho)\overline{u}_h,\\ 
	\overline{u^\prime u^\prime}(z) &=& \rho\overline{(u^\prime u^\prime)}_l + 
(1-\rho)\overline{(u^\prime u^\prime)}_h - 2\rho(1-\rho)\overline{u}_l 
\overline{u}_h,
\end{subeqnarray}
\returnthesubequation
where,
\begin{subeqnarray}\label{eqn:LocalFluctuations}
	\gdef\thesubequation{\theequation \mbox{\textit{a}},\textit{b}}
	\overline{(u^\prime u^\prime)}_l \equiv \overline{u_l^2} - 
\rho(\overline{u}_l)^2 \quad\text{and}\quad
	\overline{(u^\prime u^\prime)}_h \equiv \overline{u_h^2} - 
(1-\rho)(\overline{u}_h)^2.
\end{subeqnarray}
\returnthesubequation

The low-$|\Ric_f|$ mean velocity and temperature profiles are shown by the solid 
curves in figures \ref{fig:MeanVelTemp}($c$) and \ref{fig:MeanVelTemp}($d$), 
respectively, whereas the overall mean profiles are redrawn as dashed lines. The 
conditional mean profiles of the low-$|\Ric_f|$ regions are clearly different from 
the overall mean profiles of VC. The conditional mean velocity profiles (figure 
\ref{fig:MeanVelTemp}$c$) exhibit steeper slopes close to the wall and also larger 
maxima. These two observations imply that the local boundary layer in the 
low-$|\Ric_f|$ region is more vigorously animated by a turbulent wind as compared to 
the overall boundary layer. However, the conditional mean profiles still do not 
exhibit the log-profile according to (\ref{eqn:LogLaws}$a$). The absence of the 
log-profile suggests that for the present $\Ray$ range, we would not expect the 
$\Rey\sim\Ray^{1/2}$-ultimate regime scaling as predicted according to 
\cite{Grossmann+Lohse.2011}.

In contrast, the conditional mean temperature profiles (figure 
\ref{fig:MeanVelTemp}$d$) exhibit some degree of collapse at higher $\Ray$ between 
$30\lesssim z^+ \lesssim 80$ which agrees with the log-linear slope of 
(\ref{eqn:LogLaws}$b$) with $B=5.1$ (see blue straight dashed line). This collapse  
is interesting because it suggests that locally, the thermal boundary layers are 
log-dependent on the wall-normal distance as assumed by 
\cite{Grossmann+Lohse.2011,Grossmann+Lohse.2012} for turbulent thermal convection. 

With reference to figure \ref{fig:MeanVelTemp}, we emphasize that we perform a 
direct comparison with the log-laws of the wall. For the mean velocity profiles, we 
envision that the profiles will eventually exhibit the log-law once a wider 
separation of scales is achieved at very high $\Ray$. As an example of sufficient 
scale separation, we refer to the mean streamwise velocity profile of the wall-jet 
flow \citep{Wygnanski+Katz+Horev.1992,Eriksson+Others.1998}.
The mean velocity profiles of the wall-jet flow are statistically similar to the 
mean streamwise velocity profile of VC, but have been shown to exhibit the log-law 
of the wall. As a comparison, we include a candidate mean streamwise velocity 
profile from the wall-jet flow of \cite{Eriksson+Others.1998} in figure 
\ref{fig:MeanVelTemp}($c$), see red curve with open symbols. The wall-jet profile 
exhibits a higher maxima compared to the VC profile at $\Ray=10^9$ and agrees well 
with the log-law of the wall. Therefore, from the perspective of classical 
dimensional arguments of the `inner' and `outer layer' \citep[][Chapter 
7]{Pope2000turbulent}, we infer that the outer layer of VC corresponds to the 
location near the maxima of the streamwise velocity, as opposed to the region close 
to the channel-centre. The log-profiles in VC are also expected since similar trends 
have also been measured for RBC and Taylor--Couette flows, but in different regions 
of the flow. In the ultimate regime of RBC, the temperature log-profile was 
found in the bulk flow region \citep{Ahlers+Others+Log.2012,Wei+Ahlers.2014}, 
whereas in the classical regime of RBC, the log-profile was found in 
high-$|\Ric_f|$ regions where thermal plumes are emitted 
\citep{vanderPoel+coworkers.2015.logarithmic}. In Taylor--Couette flows, a 
log-profile was found in the azimuthal velocity profile in plume emission regions, 
as well as in the (appropriately shifted) overall angular velocity profile 
\citep{Ostilla+Coworkers.2014}.

For comparison, we also include the conditional mean profiles of the high-$|\Ric_f|$ 
regions in figures \ref{fig:MeanVelTemp}($e$) and \ref{fig:MeanVelTemp}($f$). The 
mean profiles are lower in magnitude than the overall profiles, which can be 
attributed to a weaker influence of the turbulent wind in the high-$|\Ric_f|$ 
regions. In addition, the mean temperature profiles exhibit some degree of collapse 
but does not fit the log-linear trend of (\ref{eqn:LogLaws}$b$).

\section{Scaling of local $\Nus$ and $\Rey$}\label{sec:NuReScaling}
\begin{figure}
	\centering
	\centerline{\includegraphics{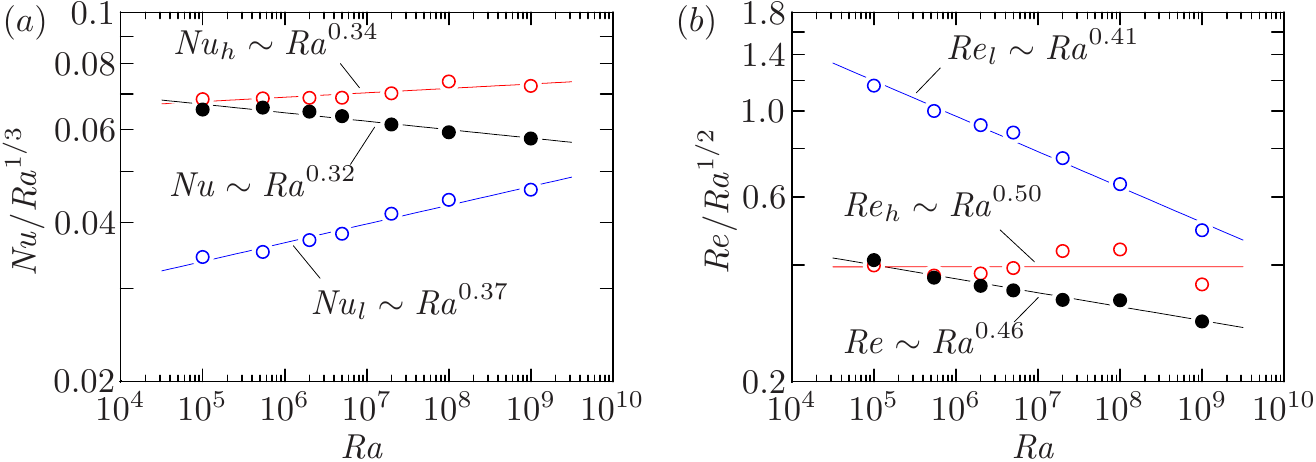}}
	\vspace*{1em}
	\caption{\label{fig:NuReScaling}Compensated plots for ($a$) $\Nus$ and ($b$) 
	$\Rey$. Black symbols denote the overall $\Nus$ and $\Rey$ results, blue symbols 
	denote the higher-shear, low-$|\Ric_f|$ results and red symbols denote the 
	lower-shear, high-$|\Ric_f|$  results. All curve-fits are from a least-squares 
	fit to an effective power law.}
\end{figure}
We can further test the characteristics of the low-/high-$|\Ric_f|$ regions in VC by 
computing the local Nusselt and Reynolds number trends in the corresponding regions. 
Within the low-$|\Ric_f|$ regions, the local Nusselt number and Reynolds number are 
defined respectively as
\begin{subeqnarray} \label{eqn:NuReLowRi}
	\gdef\thesubequation{\theequation \mbox{\textit{a}},\textit{b}}
	\Nus_l \equiv	\dfrac{f_{w,l}H}{\Delta \kappa}\quad\text{and}\quad
	\Rey_l \equiv \dfrac{U_{rms,l} H}{\nu},
\end{subeqnarray}
\returnthesubequation
where $f_{w,l} \equiv -\kappa (\romd \overline{\varTheta}_l/\romd z)|_w$. To define 
the local Reynolds number, we define $U_{rms,l}$ as the local turbulent wind and 
adopt (\ref{eqn:LocalFluctuations}) to compute the local velocity fluctuations. 
Therefore, we define $U_{rms,l} \equiv \left[\overline{(u^\prime 
u^\prime)}_l\right]^{1/2}$ evaluated at the wall-normal location $z = 
\delta_{max,l}$, where $\delta_{max,l}$ is the wall-normal distance to the maximum 
of $\overline{u}_l$. The local Nusselt and Reynolds numbers in the high-$|\Ric_f|$ 
regions follow the definition in (\ref{eqn:NuReLowRi}) with the corresponding 
high-$|\Ric_f|$ quantities. The overall Nusselt and Reynolds numbers are defined 
using the overall $f_w$ and $U_{rms}$. The results of the scaling of the Nusselt and 
Reynolds numbers are shown in figure \ref{fig:NuReScaling}.

In figure \ref{fig:NuReScaling}, we emphasise the scaling of the Nusselt and 
Reynolds numbers by plotting the trends in compensated forms, \ie\,$\Nus/\Ray^{1/3}$ 
versus $\Ray$ and $\Rey/\Ray^{1/2}$ versus $\Ray$. The overall $\Nus$ and $\Rey$ are 
denoted by solid black symbols, $\Nus_l$ and $\Rey_l$ by open blue circles, and 
$\Nus_h$ and $\Rey_h$ by open red circles. Starting with the scaling of the local 
Nusselt number (figure \ref{fig:NuReScaling}$a$), we find that $\Nus_l \sim 
\Ray^{0.37}$ from a least-squares fit to an effective power law, which is steeper
than the relation $\Nus \sim \Ray^{0.32}$ for the overall Nusselt number 
\citep{Ng+Ooi+Lohse+Chung.2014}. Interestingly, the effective scaling exponent 
computed for the low-$|\Ric_f|$ regions, which is $0.37$, is close to the exponent 
reported by \cite{He+others.2012} in their experiments for RBC at large $\Ray$, 
which is $0.38$. This result appears consistent with the predictions of the Nusselt 
number scaling relation of a $1/2$-power law scaling with logarithmic corrections in 
the ultimate regime \citep{Grossmann+Lohse.2011}. For VC, the logarithmic 
corrections for the effective scaling of $\Nus_l$ are associated with the log-linear 
slope of the conditional mean temperature profile in the high-shear, low-$|\Ric_f|$ 
regions (see figure \ref{fig:MeanVelTemp}$d$). On the other hand, the scaling of 
$\Nus_h$ follows $\Nus_h\sim\Ray^{0.34}$ which is steeper than the scaling for 
overall $\Nus$ but is still less steep compared to the scaling of $\Nus_l$.

We note that the local scaling of the Nusselt number is sensitive to the upper 
limit of the $|\Ric_f|$ criterion established in \S\,\ref{sec:RichardsonNo}. For a 
higher limit, the conditional mean statistics of the higher-shear regions are 
contaminated by the laminar-like lower-shear regions (see Appendix 
\ref{sec:AppendixA}) and as a result, the effective $\Ray$-scaling exponent of the 
local Nusselt number is closer to the global $0.31$ exponent. For example, the 
effective $\Ray$-scaling exponent for $\Nus_l$ is approximately $0.33$ using $0 < 
|\Ric_f| \leqslant 0.5$ and is approximately $0.39$ using $0 < |\Ric_f| \leqslant 
0.05$. In the former, the scaling exponent of $0.33$ is closer to $0.31$, whereas in 
the latter, the scaling exponent of $0.39$ is clearly steeper than $0.31$ and is 
consistent with the scaling exponent reported by \cite{He+others.2012}.

For the scaling of the Reynolds numbers (figure \ref{fig:NuReScaling}$b$), we find 
that $\Rey_l\sim\Ray^{0.41}$, $\Rey_h\sim\Ray^{0.50}$ and $\Rey\sim\Ray^{0.46}$. 
Based on the much lower effective scaling exponent for $\Rey_l$, it is clear that 
the Reynolds number in the high-shear regions do not exhibit the $1/2$-power-law 
ultimate regime scaling as predicted in GL theory \citep{Grossmann+Lohse.2011}. 
However, this result is not surprising since the conditional mean velocity profiles 
for the low-$|\Ric_f|$ regions (see figure \ref{fig:MeanVelTemp}$c$) do not exhibit 
the log-linear profile of  (\ref{eqn:LogLaws}$a$), which is assumed for fully 
turbulent thermal convection \citep{Grossmann+Lohse.2011} and is a key assumption of 
the GL theory for the $\Rey$-scaling in the ultimate regime.

\begin{figure}
	\centering
	\centerline{\includegraphics{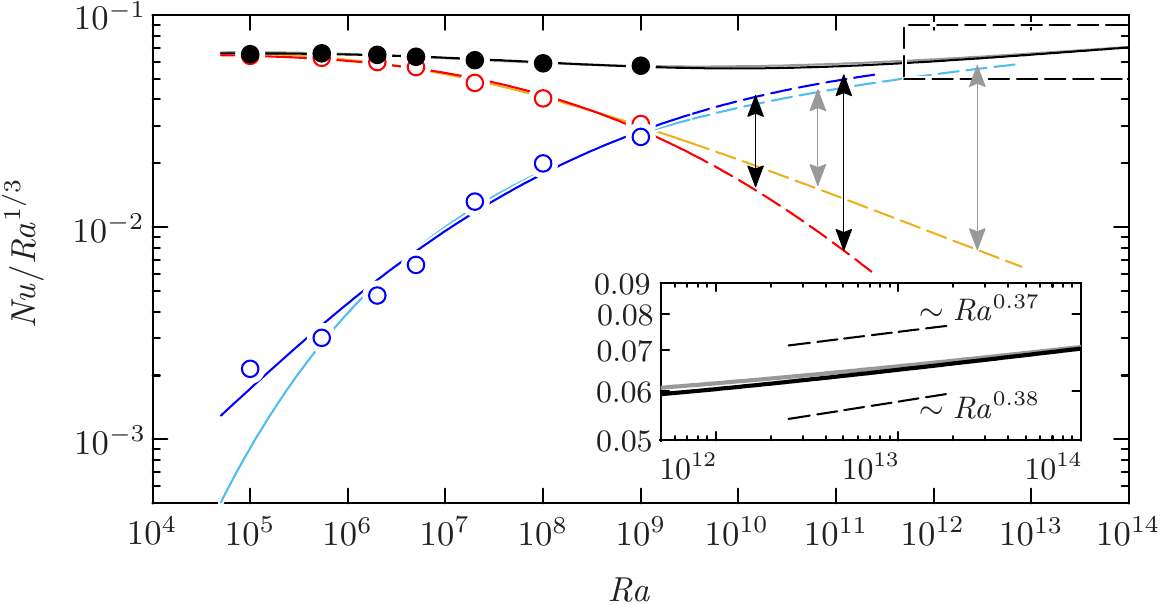}}
	\vspace*{1em}
	\caption{\label{fig:StreakyNuDominates}Relative contributions of 
	the local Nusselt numbers with increasing $\Ray$. The solid black symbols denote 
	the overall $\Nus$, the blue symbols denote $\rho\Nus_l$ and the red symbols 
	denote $(1-\rho)\Nus_h$. The blue and red curves represents 
	$\rho_{fit,1}\Ray^{0.37}$ and $(1-\rho_{fit,2})\Ray^{0.34}$, respectively, and 
	the sum equals to the black curve. The cyan and orange curves represent 
	$\rho_{fit,2}\Ray^{0.37}$ and $(1-\rho_{fit,2})\Ray^{0.34}$, respectively, and 
	the sum equals to the grey curve. The inset plot compares the extrapolated 
	curves in the dashed box with the $\Ray$-scaling exponent of 0.37 and 0.38. The 
	trends indicate that the contributions from $\Nus_l$ dominate the contributions 
	from $\Nus_h$ with increasing $\Ray$.}
\end{figure}

Based on our results in figure \ref{fig:NuReScaling}($a$) for the effective scaling 
exponents of $\Nus_l$ and $\Nus_h$, we estimate the relative contributions of 
$\Nus_l$ and $\Nus_h$ by making use of the relation
\begin{equation}
\Nus = \rho \Nus_l + (1-\rho) \Nus_h, \label{eqn:NuSplit}
\end{equation}
which can be derived using the conditional averaging procedure described by 
(\ref{eqn:ConditionalAveProcess}). We then model the two terms on the 
right-hand-side of (\ref{eqn:NuSplit}) using both curve-fits 
(\ref{eqn:TurbFractionCompositeFit}) and (\ref{eqn:TurbFractionGompertzFit}), and 
the effective power-law curve-fits for $\Nus_l$ and $\Nus_h$ from figure 
\ref{fig:NuReScaling}($a$), thus,
\begin{subeqnarray}\label{eqn:NuSplitModel}
	\gdef\thesubequation{\theequation \mbox{\textit{a}},\textit{b}}
	\rho \Nus_l \sim \rho_{fit}\Ray^{0.37}\quad\text{and}\quad (1-\rho)\Nus_h \sim 
	(1-\rho_{fit})\Ray^{0.34}.
\end{subeqnarray}
\returnthesubequation
The two modelled terms on the right-hand-side of (\ref{eqn:NuSplitModel}) are 	
plotted in figure \ref{fig:StreakyNuDominates}, where the blue-red lines correspond 
to $\rho_{fit,1}$ and the cyan-orange lines correspond to $\rho_{fit,2}$. The blue 
and red symbols in figure \ref{fig:StreakyNuDominates} correspond to the results 
from our DNS, \ie\,the left-hand-side of (\ref{eqn:NuSplitModel}), respectively. The 
solid black symbols are the overall $\Nus$-values, which are exactly equal to the 
sum of the values denoted by the blue and red symbols.

From figure \ref{fig:StreakyNuDominates}, we find that the contribution of $\rho 
\Nus_l$ increases with increasing $\Ray$, as expected. When we extrapolate the two 
modelled terms in (\ref{eqn:NuSplitModel}), shown by the dashed lines in figure 
\ref{fig:StreakyNuDominates}, the contribution from $\Nus_l$ dominates the 
contribution from $\Nus_h$ at higher $\Ray$. For instance, the ratio of $\rho 
\Nus_l$ to $(1-\rho)\Nus_h$ for $\rho_{fit,1}$ ranges from approximately 3 to 7 
between $\Ray\approx 1.5\times 10^{10}$ to $1.2\times 10^{11}$ (black arrows in 
figure \ref{fig:StreakyNuDominates}) and, coincidently for 
$\rho_{fit,2}$, also ranges from approximately 3 to 7 between $\Ray \approx 
6.5\times 10^{10}$ to $2.8\times 10^{12}$ (grey arrows in figure 
\ref{fig:StreakyNuDominates}). The two pairs of $\Ray$-values 
correspond to the estimated $80\%$ to $90\%$ of the near-wall area that are 
presumably covered by the streaky, higher-shear regions 
(\cf\,\S\,\ref{sec:TurbFraction}). For the sake of obtaining estimates, we further 
extrapolate the sum of the modelled terms from (\ref{eqn:NuSplitModel}) and the 
trends are shown as solid black and solid grey curves in figure 
\ref{fig:StreakyNuDominates}. At $\Ray \gtrsim 10^{13}$, the slope of the two 
extrapolated trends become comparable to the $\Ray$-scaling exponent of between 
0.37-0.38 (see inset of figure \ref{fig:StreakyNuDominates}), which is expected 
since the relative contribution of $(1-\rho)\Nus_h$ diminishes at very high-$\Ray$.

For standard RBC, \cite{Shang+Tong+Xia.2008} performed a similar decomposition of 
the heat transfer, namely in a plume-dominated part $\Nus_{pl}\sim\Ray^{0.24}$ and a 
bulk-dominated part $\Nus\sim\Ray^{0.49}$, with the bulk part gaining dominance 
beyond $\Ray_c \approx 10^{14}$ \citep[see 
also][]{Shang+others.2003,Grossmann+Lohse.2004}. The much lower $\Ray$-values for 
when the ultimate regime scaling dominates in VC as compared to standard RB 
(where it is $\Ray_c \approx 10^{14}$) may be associated with the more efficient 
driving in the VC configuration since gravity is parallel to the walls in VC, 
whereas gravity is orthogonal to the walls in RBC. For the shear-driven 
Taylor--Couette flow, the driving is even more efficient and the transition to the 
ultimate regime occurs at the Taylor number $\Tay_c \simeq 3\times 10^8$ for an 
inner-to-outer radius ratio of 0.71 \citep{Grossmann+Lohse+Sun.2016}.

\begin{figure}
	\centering
	\centerline{\includegraphics{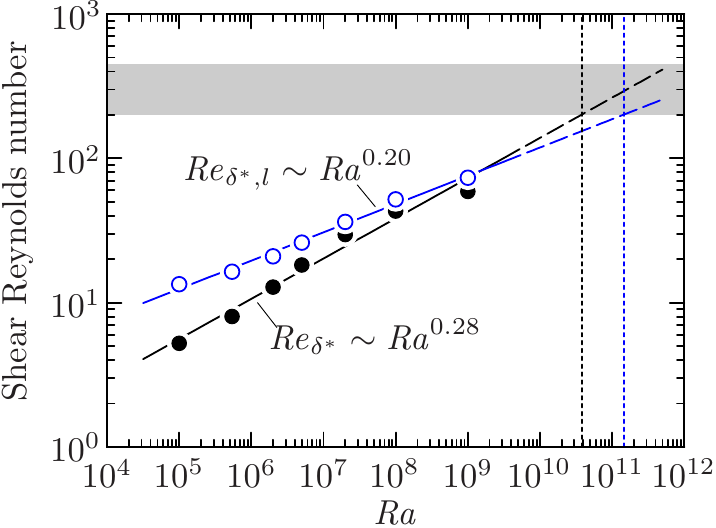}}
	\vspace*{1em}
	\caption{\label{fig:ShearRe}Plots of shear Reynolds number versus $\Ray$. The 
	blue circles denote the low-$|\Ric_f|$ shear Reynolds number 
	($\Rey_{\delta^\ast,l}$) whereas the solid black circles denote the overall 
	shear Reynolds number ($\Rey_{\delta^\ast}$). The grey region indicates the 
	range of the shear Reynolds number between 200 and 450. At the shear Reynolds 
	number value of 200, the $\Ray$-value is estimated to be $1.5\times 10^{11}$ for 
	the low-$|\Ric_f|$ shear Reynolds number	and $3.8\times 10^{10}$ for the 
	overall shear Reynolds number (demarcated by the dotted lines).}
\end{figure}
In figure \ref{fig:ShearRe}, we compute the shear Reynolds number based on the 
displacement thickness $\delta^\ast$ according to the definition $\Rey_{\delta^\ast} 
\equiv U\delta^\ast/\nu$ for the low-$|\Ric_f|$ and overall regions. For the 
low-$|\Ric_f|$ regions, we define $U\equiv\text{max}(\overline{u}_l)$ and
$\delta^\ast \equiv \int_{0}^{\delta_{max,l}} 
(1-\overline{u}_l/\text{max}(\overline{u}_l))\romd z$. For the overall shear 
Reynolds number, we define $U\equiv\text{max}(\overline{u})$ and $\delta^\ast \equiv 
\int_{0}^{\delta_{max}} (1-\overline{u}/\text{max}(\overline{u}))\romd z$. At some 
critical value, the shear Reynolds number can provide an indication of the 
laminar-to-turbulent transition \citep[for example, 
$(\Rey_{\delta^\ast})_{cr}\approx 420$][]{Landau+Lifshitz.1987}. However, since the 
theoretical value for the critical shear Reynolds number is presently unknown for 
VC, we choose a range of $\Rey_{\delta^\ast}$ from 200 to 450 as an approximate 
indicator for the transition to turbulence in VC. This range is shown as the grey
region in figure \ref{fig:ShearRe}. The results from the low-$|\Ric_f|$ regions are 
denoted by the open blue circles and the results from the overall regions are 
denoted by the solid black circles. The associated effective power-law fits are 
$\Rey_{\delta^\ast,l}\approx 1.3\Ray^{0.20}$ (blue line) and 
$\Rey_{\delta^\ast}\approx 0.23\Ray^{0.28}$ (black line). 

From figure \ref{fig:ShearRe}, we find that $\Rey_{\delta^\ast,l}$ is larger than 
$\Rey_{\delta^\ast}$ for the present $\Ray$-range of our DNS dataset. By 
extrapolating the power-law fits (dashed lines in figure \ref{fig:ShearRe}), we find 
that at the shear Reynolds number value of 200, the $\Ray$-value is estimated to
be $\approx 3.8\times 10^{10}$ for the overall region and $\approx 1.5\times 
10^{11}$ for the low-$|\Ric_f|$ region. Both $\Ray$-values appear consistent with 
the $\Ray$-values estimated when the values of the turbulent fraction $\rho = 0.8$ 
and $0.9$ (see figure \ref{fig:TurbFraction}). Conservatively, our results suggest 
that we could expect the ultimate regime in VC to occur at $\Ray \gtrsim 10^{11}$. 
Physically, the near-wall region is expected to be dominated by streaky, high-shear 
regions, however, further investigations at much higher $\Ray$ are necessary in 
order to verify this hypothesis.

\section{Conditional spectra}\label{sec:CondSpectra}
\begin{figure}
	\centering
	\centerline{\includegraphics{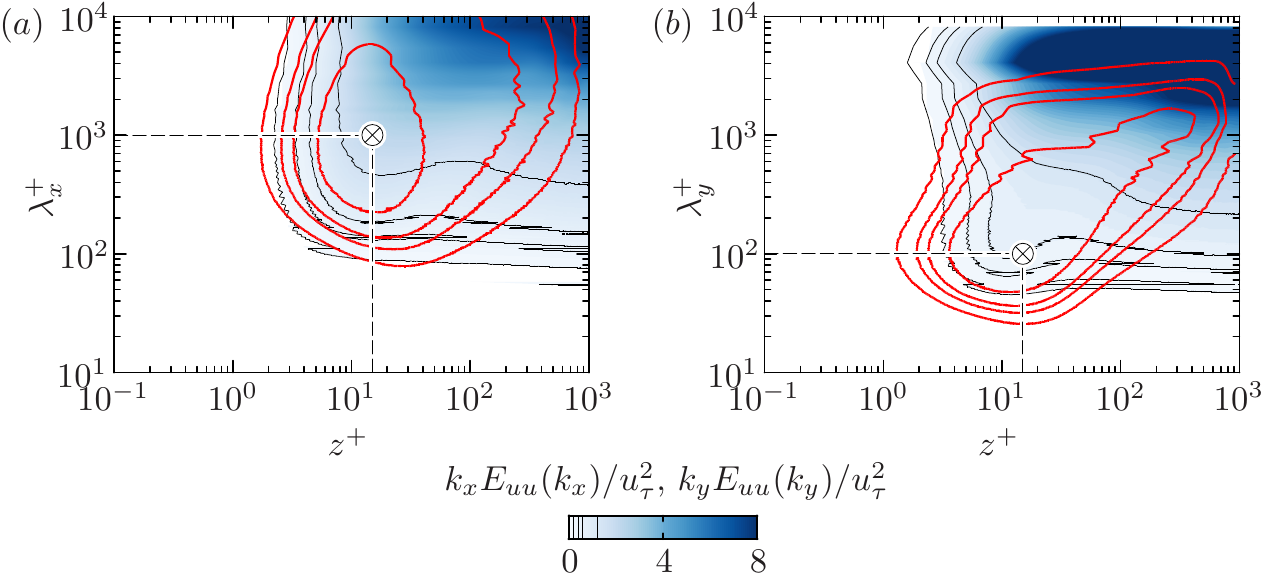}}
	\caption{\label{fig:FullSpec}One-dimensional premultiplied overall spectra of 
	$u$ for $\Ray=10^9$ shown as blue contours: ($a$) streamwise direction and ($b$) 
	spanwise direction. The red contours are the spectra from DNS of channel flow at 
	$\Rey_\tau \approx 934$		
	\citep[\cf][]{delAlamo+Jimenez+Zandonade+Moser.2004}. For both VC and channel 
	flow spectra, the contour levels are fixed at $0.2$ (outermost), $0.4$, $0.6$ 
	and $1.2$ (indicated by the thinner lines in the colorbar). The $\otimes$ 
	symbols denote the inner sites of channel flow spectra, which are respectively 
	at ($z_p^+$,$\lambda_{x,p}^+$) $\approx$ ($15$,$1000$) and		
	($z_p^+$,$\lambda_{y,p}^+$) $\approx$ ($15$,$100$).}
\end{figure}

In this section, we investigate the similarities between the near-wall structures in 
VC and in a canonical wall-bounded flow setup. In particular, we look for evidence 
of buffer-region features in the near-wall region of VC, which would suggest that 
the VC flow is transitioning to the ultimate regime with log-variations in the mean 
profile. Our basis of comparison is the unique structural features in the buffer 
region of the flow at $z^+ \approx 15$, \ie\,before the start of the log-law of the 
wall. The features in this region are commonly referred to as near-wall streaks 	
\citep{Kline+Others.1967,Hutchins+Marusic_MeanderingFeatures.2007,Marusic+Others.2010}.
In canonical wall-bounded flows, the near-wall streaks are robust features of the 
buffer region: they are found at every location in the wall-parallel plane $z^+ 
\approx 15$, contribute to the peak in streamwise velocity fluctuations and exhibit 
a characteristic spanwise spacing of $\lambda_{y}^+ \approx 100$. Here, $\lambda_y = 
2\pi k_y^{-1}$ the spanwise wavelength. In the following, we conduct a detailed 
comparison between VC and canonical channel flow with the aid of the one-dimensional 
spectra.
	
We define the one-dimensional energy spectra of streamwise velocity $u$ as 
$E_{uu}(k_j)$, where $(u^{\prime})^2 = 2\int_0^\infty E_{uu}(k_j)\,\romd{k_j}$, 
$k_j$ is the wavenumber in the $j$-th direction and $j=1,2$. Thus, $k_1=k_x$ the 
streamwise wavenumber and $k_2=k_y$ the spanwise wavenumber. As a reference, we 
first plot the overall spectra of $u$ for VC at $\Ray=10^9$ in figure 
\ref{fig:FullSpec}, shown as filled blue contours. Following 
\cite{Hutchins+Marusic_MeanderingFeatures.2007}, we plot the energy spectra at all 
wall-normal locations in premultiplied form, \ie $k_xE_{uu}(k_x)$ versus $\lambda_x 
= 2\pi k_x^{-1}$ the streamwise wavelength (figure \ref{fig:FullSpec}$a$) and 
$k_yE_{uu}(k_y)$ versus $\lambda_y = 2\pi k_y^{-1}$ the spanwise wavelength (figure  
\ref{fig:FullSpec}$b$). In premultiplied form and on a logarithmic plot, the 
contours of the energy spectra visually represents the distribution of energy that 
resides at the corresponding wavelengths (similar to the representation of 
$|\Ric_f|$-distribution in figure \ref{fig:RiPDF}$b$). The plotted spectra are 
normalised by $u_\tau^2$.

In figure \ref{fig:FullSpec}, the contours of the spectra for VC appear unclosed  
and this is because the peak of the spectra occurs at the peak of $u_{rms}$, which 
is at the channel  centre \citep[see for example figure 3($a$) 
in][]{Ng+Ooi+Lohse+Chung.2014}. In addition, the  unclosed spectra indicate that the 
longest energy containing wavelengths in VC exceed the  streamwise and spanwise 
domain lengths of the present setup at $\Ray=10^9$. Therefore, for higher $\Ray$, it 
would  be necessary to conduct simulations in the periodic-domain sizes that are 
larger than $L_x\times  L_y = 8H\times 4H$ in each direction if a higher degree of 
convergence is desired. Nevertheless, for  the subsequent analysis into structure of 
the streaky regions, the choice of a larger periodic-domain size  would 
qualitatively give the same results.

To compare the distribution of spectra for VC, we include the spectra from DNS of 
channel flow at the friction Reynolds number $\Rey_\tau \equiv u_\tau\delta/\nu 
\approx 934$ \citep[\cf][]{delAlamo+Jimenez+Zandonade+Moser.2004}, where $\delta$ 
is the channel half-width, in figure \ref{fig:FullSpec}. The channel flow spectra 
are shown as red contours. In contrast to VC, the spectra of channel flow exhibit 
peaks in the near-wall region at $(z^+_p,\lambda_{x,p}^+) \approx (15,1000)$
and $(z^+_p,\lambda_{y,p}^+) \approx (15,100)$. These intense near-wall energies in 
channel flow correspond to the `inner' sites and are well-known characteristics of 
the near-wall streaks
\citep[\cf][]{Hutchins+Marusic_MeanderingFeatures.2007,Hutchins+Marusic_NearWallCycle.2007}.
For reference, we mark the inner sites by the symbol $\otimes$ in figure 
\ref{fig:FullSpec}. When we compare the spectra of VC and channel flow, we find 
little resemblance between the shape of the streamwise spectra of VC and channel 
flow. However, the shape of the spanwise spectra of VC and channel flow exhibit some 
similarities in the near-wall region: the contour of the channel flow spectra at 
$k_yE_{uu}(k_y)/u_\tau^2 = 1.2$ envelops the contour of the spectra for VC at 
$k_yE_{uu}(k_y)/u_\tau^2 = 0.2$. This similarity between the spanwise spectra in 
the near-wall region suggests that it may be possible to distinguish near-wall 
regions that exhibit characteristics similar to canonical wall-bounded turbulence. 
Thus, our aim is to analyse the local spectra of $u$ in the low- and high-$|\Ric_f|$ 
regions (\cf\,\S\,\ref{sec:RichardsonNo}) that exhibit the streaky and non-streaky 
regions, respectively.

\begin{figure}
	\centering
	\centerline{\includegraphics{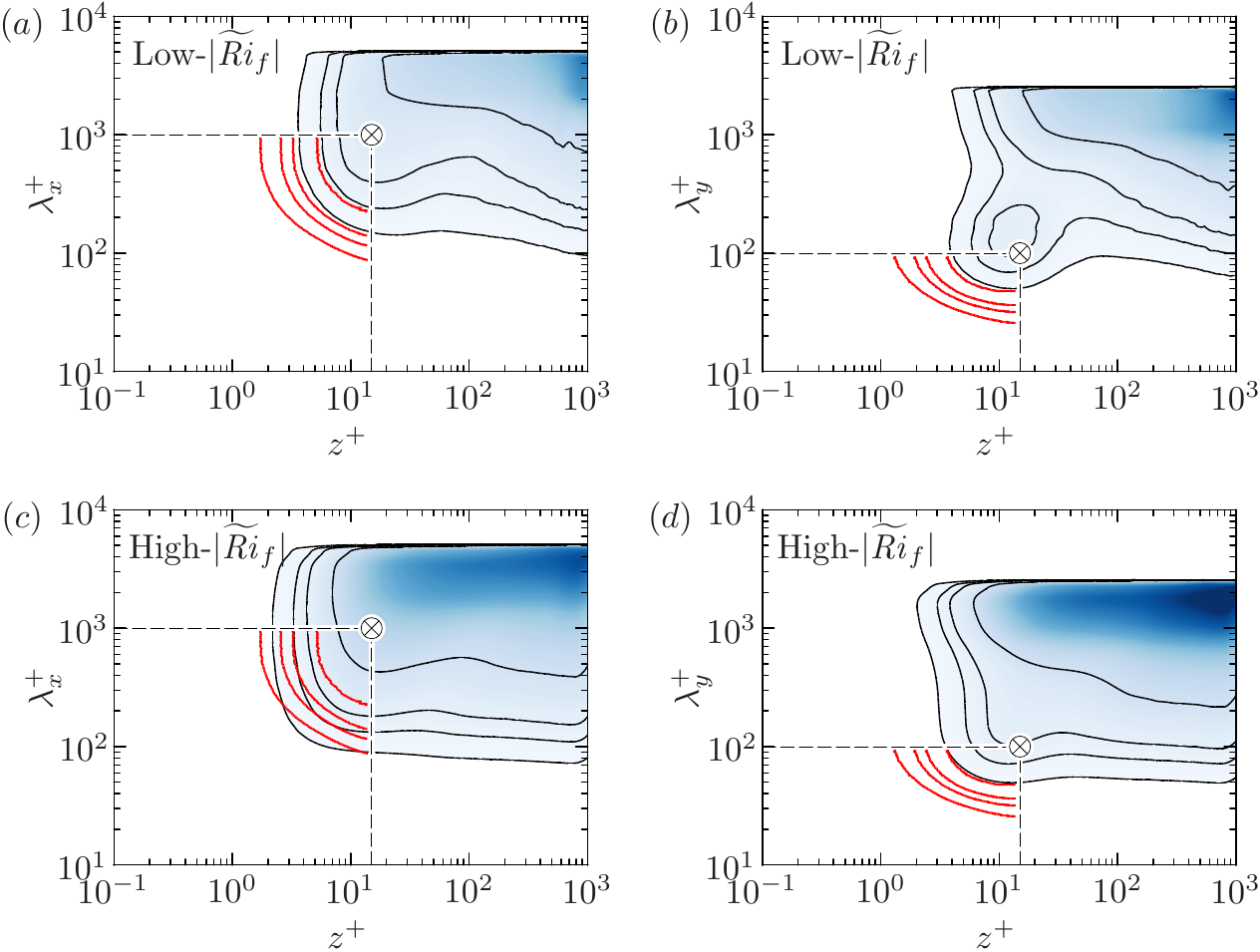}}
	\caption{\label{fig:CondSpec}One-dimensional premultiplied conditional spectra 
	of $u$ for $\Ray=10^9$ shown as blue contours: ($a$,$c$) streamwise direction 
	and ($b$,$d$) spanwise directions. ($a$,$b$) Low-$|\widetilde{\Ric}_f|$ ($0 < 
	|\widetilde{\Ric}_f| \leqslant 0.1$) spectra for higher-shear			
	regions. ($c$,$d$) High-$|\widetilde{\Ric}_f|$ ($|\widetilde{\Ric}_f| > 0.1$) 
	spectra for lower-shear	regions. The red contours are spectra from DNS of 
	channel flow at $\Rey_\tau \approx 934$			
	\citep[\cf][]{delAlamo+Jimenez+Zandonade+Moser.2004}, where only the contour 
	levels in the near-wall region and shorter wavelengths are shown for clarity. 
	The colormaps and contour levels are the same as figure \ref{fig:FullSpec}. The 
	$\otimes$ symbols denote the inner sites of channel flow spectra, which are 
	respectively at ($z^+_p$,$\lambda_{x,p}^+$) $\approx$ ($15$,$1000$) and 
	($z^+_p$,$\lambda_{y,p}^+$) $\approx$ ($15$,$100$).}
\end{figure}

In order to obtain the local spectra of $u$, we first need to define the 
domain-lengths of a local streamwise-spanwise patch. In doing so, we restrict our 
analysis of the local spectra of $u$ to a patch that is associated with either low- 
or high-$|\Ric_f|$. As a starting point, we choose the domain-lengths of the local 
patch as $2H \times H$ in the streamwise and spanwise directions. A patch is defined 
as locally shear-dominated if the plane-averaged value of $|\Ric_f|$ in the local
patch, which we denote by $|\widetilde{\Ric}_f|$, is between $0$ and $0.1$, thus 
maintaining consistency with our criterion defined in \S\,\ref{sec:RichardsonNo}. 
From this definition, we obtain two conditionally averaged spectra for each 
direction: one for a (patch-averaged) higher-shear region, 
\ie\,low-$|\widetilde{\Ric}_f|$, and another for a (patch-averaged) lower-shear 
region, \ie\,high-$|\widetilde{\Ric}_f|$.

Next, since the local $u$ is not spatially periodic in the $x$- and $y$-directions, 
we introduce periodicity by appling a normalized Hann window $W(x_i) = 
\sqrt{2/3}\,[1-\cos(2\pi n_i/N_{x_i})]$ \citep[\cf][]{Chung+McKeon.2010} in the $x$- 
and $y$-directions of the patch. This step is necessary in order to minimise 
spectral leakage when computing the spectra of a non-periodic local signal of $u$.
Here, $N_{x_i}$ is the number of grid points in the $x_i$ direction and 
$n_i=1,\ldots,N_{x_i}$. This procedure is repeated at all wall-normal locations of 
the patch. The computed local spectra are then normalised by the local 
patch-averaged friction velocity $\widetilde{u}_{\tau}$ before averaging. Figure 
\ref{fig:CondSpec} shows the results for the conditionally averaged spectra at 
$\Ray=10^9$, normalised by $\widetilde{u}_{\tau}^2$. For clarity, we compare our 
results with the channel flow spectra for $z^+\leqslant15$, $\lambda_{x}^+\leqslant 
1000$ and $\lambda_{y}^+\leqslant 100$.

Interestingly, when we closely inspect the contours of the 
low-$|\widetilde{\Ric}_f|$ spectra of VC (figure \ref{fig:CondSpec}$a$,$b$), we find 
that the shape of both streamwise and spanwise spectra have a relatively higher 
degree of resemblance with the spectra of channel flow compared to the contours in 
figure \ref{fig:FullSpec}. The magnitude of the spectra of VC remains lower than
the magnitude of the channel flow spectra. We also note the presence of a prominent 
near-wall peak in the spanwise conditional spectra of VC occurring at 
$z_{p,VC}^+\approx 15$ and $\lambda_{y,p,VC}^+ \approx 150$ (see figure 
\ref{fig:CondSpec}$b$), which is close to the inner site of the channel flow spectra.
Additionally, the value of the peak spanwise wavelength in VC is consistent with
the spanwise spacing of the streaks of approximately $100$ to $200$ viscous units 
observed in figure \ref{fig:VelTempSliceZoom}. These notable similarities between VC 
and turbulent channel flow imply that some form of the near-wall streaks is also 
locally present in VC, but the intensities of the streaks are dominated by the 
energy of considerably larger scales. From our dataset, we find that the near-wall 
similarities (both in spectra and flow visualisations) emerge at $\Ray \gtrsim 
2\times 10^7$. This suggests that for the present $\Ray$-range, the VC flow is 
undergoing a transition to a regime where the boundary layers are increasingly 
influenced by stronger shear, which is consistent with our findings in the preceding 
sections. 

For the conditionally averaged high-$|\widetilde{\Ric}_f|$ spectra of VC (figure 
\ref{fig:CondSpec}$c$,$d$), the shape of the contours retain characteristics that 
are similar to the unconditioned spectra of VC in figure \ref{fig:FullSpec}. Namely, 
the shape of the streamwise spectra of VC differs from the streamwise spectra of 
channel flow whereas the spanwise contours retain the resemblance in shape but 
without a near-wall peak.

\section{Summary and conclusion}\label{sec:Conclusion}

The boundary layers in VC with $\Ray$ ranging between $10^5$ and $10^9$ and 
$\Pran$-value of $0.709$ are found to exhibit characteristics of a transition from a 
laminar-like scaling to a turbulent ultimate-regime-type behaviour. 

Visually, the near-wall flow shows the emergence of two distinct regions with 
increasing $\Ray$ (figure \ref{fig:VelTempSlice}). The first region appears streaky 
and correlates with higher-shear regions, whereas the second region is not streaky 
and correlates with lower-shear regions. The two regions are well-described by a 
flux Richardson number criterion where low-$|\Ric_f|$ regions correspond to the 
streaky regions and high-$|\Ric_f|$ regions correspond to the non-streaky regions 
(figure \ref{fig:RichardsonSlice}). With increasing $\Ray$, the streaky 
low-$|\Ric_f|$ regions occupy increasingly larger fractions of the wall-area (figure 
\ref{fig:TurbFraction}). This result suggests that the transition from the
classical regime to the ultimate regime in VC manifests in the boundary layers as 
increasingly larger wall-area coverage of shear-dominated regions. Based on the 
trend of the increasing shear-dominated wall-areas, we estimate that at $\Ray\gtrsim 
10^{11}$ the boundary layers of VC are expected to be dominated by the streaky 
higher-shear structures for the $\Pran$-value of this study.

On further analysis, we find that the local statistics in the streaky low-$|\Ric_f|$ 
regions exhibit trends that agree with turbulent boundary layer behaviour. In 
particular, the conditionally averaged mean temperature profiles of the 
low-$|\Ric_f|$ regions show good agreement with the log-linear slope of the log-law 
of the wall for mean temperature (figure \ref{fig:MeanVelTemp}$d$). No log-linear 
trends are found in the low-$|\Ric_f|$ conditional mean velocity profiles, both 
high-$|\Ric_f|$ conditional mean velocity and temperature profiles, and the overall 
unconditioned mean velocity and temperature profiles.

In the streaky low-$|\Ric_f|$ regions, the local Nusselt and Reynolds number follow 
an effective power scaling of $\Nus_l \sim \Ray^{0.37}$ and $\Rey_l \sim 
\Ray^{0.41}$ (figure \ref{fig:NuReScaling}). The effective scaling exponent of 
$0.37$ for $\Nus_l$ appears consistent with the logarithmically corrected 
$1/2$-power-law scaling predictions of the Nusselt number scaling relation in the
ultimate regime. Additionally, the contributions from the Nusselt number in 
low-$|\Ric_f|$ regions increasingly dominate the contributions from the 
high-$|\Ric_f|$ regions (figure \ref{fig:StreakyNuDominates}), which suggests that 
at sufficiently high $\Ray$, the VC flow undergoes a transition to the ultimate 
regime presumably with $\Nus \sim \Ray^{0.37}$. In contrast, the effective scaling 
exponent of $0.41$ for $\Rey_l$ is lower than the $1/2$-power-law scaling 
predictions for the Reynolds number scaling relation in the ultimate regime. The 
lower exponent in the $\Rey_l$ scaling may be attributed to the local wind not being 
sufficiently strong to animate the turbulent boundary layers in the sense of Prandtl 
and von K{\'a}rm{\'a}n, as shown by the absence of a logarithmic-variation in the 
low-$|\Ric_f|$ conditional mean velocity profiles (figure \ref{fig:MeanVelTemp}$c$). 
Indeed, investigations at much higher $\Ray$ are necessary in order to determine 
incipient logarithmic-variations in the mean velocity and temperature profiles, and 
how the local effective power-law scalings for Nusselt and Reynolds number in 
higher-shear regions are affected.

Finally, a comparison of our VC data and turbulent channel flow data at matched 
friction Reynolds number $\Rey_\tau$ reveals that the streaky regions in VC are 
reminiscent of the near-wall streaks in canonical wall-bounded flows. We find that a 
near-wall peak is present in the conditional low-$|\Ric_f|$ spectra of VC in the 
spanwise direction (figure \ref{fig:CondSpec}$b$). The corresponding wall-normal 
location is $z_{p,VC}^+\approx 15$ with a spanwise spacing of approximately 150 
viscous length scales. This near-wall peak in the spectra of VC not only resembles 
the inner site of the spectra of turbulent channel flow, but the spanwise spacing of 
150 viscous length scales is also consistent with the well-known spanwise Kline
spacings of 100 viscous length scales in turbulent wall-bounded flows (see figure 
\ref{fig:VelTempSliceZoom}$b$,$d$).

Collectively, the results of the local logarithmic-variation in the mean temperature 
profile, higher exponent scaling for the Nusselt number and characteristics of the 
near-wall streaks suggest that the boundary layer in VC is undergoing a transition 
from the classical regime to an ultimate regime for the present $\Ray$ range. This 
transition is characterised by an increasing area coverage of shear-dominated 
patches in the near-wall region with increasing $\Ray$ and are consistent with the 
recent findings for Rayleigh--B{\'e}nard convection and Taylor--Couette flows. 

\section*{Acknowledgements}
This work was supported by the resources from the National Computational 
Infrastructure (NCI) National Facility in Canberra Australia, which is supported by 
the Australian Government, and the Pawsey Supercomputing Centre, which is funded by 
the Australian Government and the Government of Western Australia. DL acknowledges 
support from FOM and the Max Planck Center for Complex Fluid Dynamics (University 
of Twente campus).

\appendix
\section{Sensitivity of $|\Ric_f|$ threshold}\label{sec:AppendixA}
\begin{figure}
	\centering
	\centerline{\includegraphics{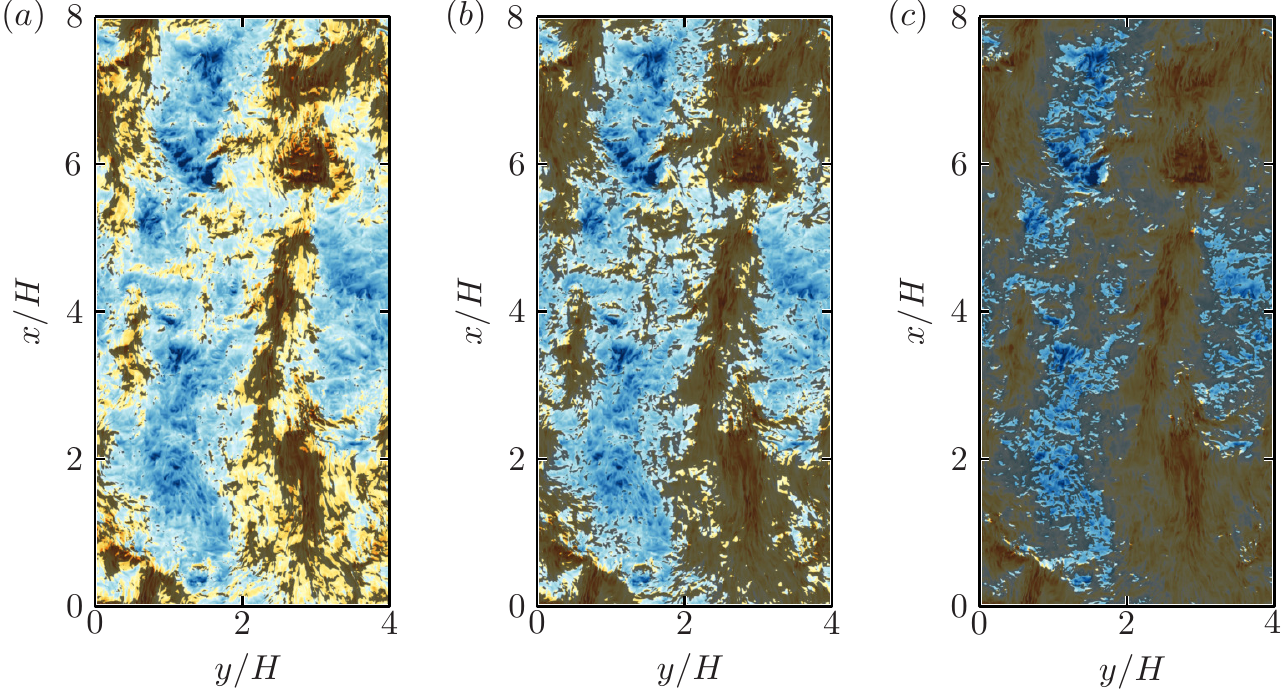}}
	\vspace*{0.5em}
	\caption{\label{fig:RifPatch}Plots of three Richardson number 
		thresholds, shown as grey contours, overlayed on wall-parallel planes of 
		streamwise velocity for $\Ray = 10^9$. The thresholds are ($a$) $0 < 
		|\Ric_f| \leqslant 0.05$, ($b$) $0 < |\Ric_f| \leqslant 0.1$ (reproduced 
		from figure \ref{fig:RichardsonSlice}$c$) and ($c$) $0 < |\Ric_f| \leqslant 
		0.5$. The velocity colormap is the same as in figure 		
		\ref{fig:VelTempSlice}.}
\end{figure}

To illustrate the sensitivity of the $|\Ric_f|$ threshold proposed in 
\S\,\ref{sec:RichardsonNo}, we plot the wall-area coverage for three threshold 
$|\Ric_f|$-values for $\Ray=10^9$ in figure \ref{fig:RifPatch}. The shaded grey 
contours are the higher-shear regions-of-interest, $\rho$, and the values are $\rho 
\approx 0.28$, $0.58$ and $0.92$ for $0 < |\Ric_f| \leqslant 0.05$, $0 < |\Ric_f| 
\leqslant 0.1$ (used in this study) and $0 < |\Ric_f| \leqslant 0.5$. Note that the 
values of $\rho$ can also be calculated by integrating the area under the 
premultiplied PDF in figure \ref{fig:RiPDF}($b$) for the corresponding 
$\Ray$-value. From figure \ref{fig:RifPatch}($c$), the grey contours encompass a 
larger area and include portions of the lower-shear regions, in blue; the local 
statistics in the higher-shear regions are therefore contaminated by laminar-like 
lower-shear regions. In contrast, the lower threshold values (figures 
\ref{fig:RifPatch}$a$ and $b$) encompass smaller areas and only parts of the 
higher-shear regions, in red.

\bibliographystyle{jfm}


\end{document}